\documentclass{aa}  
\usepackage{graphicx}
\usepackage{txfonts}
\begin{document}
   \title{GALAXY EVOLUTION FROM DEEP MULTI-WAVELENGTH INFRARED SURVEYS: A PRELUDE TO HERSCHEL
}


\titlerunning{Galaxy Evolution at IR Wavelengths}
\authorrunning{Franceschini, A., et al.}

   \author{Alberto~Franceschini\inst{1}, Giulia~Rodighiero\inst{1}, Mattia~Vaccari\inst{1},
Stefano~Berta\inst{1,2}, Lucia~Marchetti\inst{1}, Gabriele~Mainetti\inst{1}
     }

   \offprints{A. Franceschini}

   \institute{Dipartimento di Astronomia, Universita' di Padova,
I-35122 Padova, Italy\\
              \email{alberto.franceschini@unipd.it}
             \thanks{This work is based in part on observations made with the \textit{Spitzer Space
Telescope}, which is operated by the Jet Propulsion laboratory, California
Institute of Technology, under NASA contract 1407.}
         \and
         MPE
                     }

   \date{Received June 1, 2009; accepted ... }

 
  \abstract
   {Studies of the generation and assembly of stellar populations in galaxies largely benefit from far-IR observations, considering that the IR flux is a close \textit{prior} to the rate of star formation (the bulk of which happens in dust-obscured environments). At the same time, major episodes of nuclear AGN accretion are also dust-obscured and visible in the IR.
   }
   {At the end of the \textit{Spitzer} cryogenic mission and the onset of the \textit{Herschel} era, we review our current knowledge of galaxy evolution at IR wavelengths, and model it to achieve as far as a complete view of the evolution of cosmic sources. We also develop new tools for the analysis of background fluctuations to constrain source counts in regimes of high confusion, as it happens for the \textit{Herschel} sub-mm surveys.}
   {We analysed a wide variety of new data on galaxy evolution and high-redshift source populations from \textit{Spitzer} cosmological surveys, and confront them with complementary data from mm ground-based observations and constraints from the far-IR diffuse radiation, as well as preliminary results from \textit{Herschel} surveys. 
   }
   {These data confirm earlier indications about a very rapid increase in galaxy volume emissivity with redshift up to $z\simeq 1$ [$\rho(z)\propto (1+z)^{4}$], the fastest evolution rate observed for galaxies at any wavelengths.
The observed \textit{Spitzer} counts require a combination of fast evolution for the dominant population and a bumpy spectrum with substantial PAH emission at $z\sim 1$ to 2. Number counts at long wavelengths (70 through 1100 $\mu$m) confirm these results.
All the present data require that the fast observed evolution from z=0 to 1 flattens around redshift 1 and then keeps approximately constant up to $z\simeq 2.5$ at least.  Our estimated redshift-dependent bolometric comoving energy density keeps lower at $z\gtrsim 1.5$ than some previously published results based on either large extinction corrections, or large spectral extrapolations.
}
{
The present-day IR/sub-mm data provide evidence of a distinct population of very luminous galaxies becoming dominant at $z>1$. Their cosmological evolution, peaking around $z\simeq 2$, shows a faster decay with cosmic time than lower luminosity systems, whose maximal activity is set around $z\simeq 1$, then supporting an earlier phase of formation for the most luminous and massive galaxies. 
From a comparison of our results on the comoving IR emissivity with recent estimates of the redshift-dependent stellar mass functions of galaxies, we find that the two agree well with each other if we assume standard recipes for star formation (a universal Salpeter IMF) and standard fractions ($\sim 20-30\%$) for the contribution of obscured AGN accretion.
Systematic exploitation of the forthcoming \textit{Herschel} survey data will be important for confirming all this.
}


   \maketitle
%

\section{Introduction}

Galaxy evolution has been studied in detail and is currently quite precisely known between redshift 0 and approximately 1, mostly based on deep optical imaging and spectroscopy (see, among others projects, DEEP-2, Faber et al. 2007; COMBO-17, Bell et al. 2004; K20, Cimatti et al. 2002; COSMOS, Scoville et al. 2007; VVDS, Le Fevre et al. 2005). 
One important aspect of these investigations concerns the fraction of optical-UV radiant energy emitted by young stellar populations, which is absorbed by dust in the ISM medium and re-emitted at long wavelengths. It has been shown that, while being modest for evolved galaxies like early-type spirals, and negligible in E/S0 galaxies, this fractional energy may become strong or even dominant during active phases of star formation, galaxy assembly, and nuclear activity (Kormendy and Sanders 1992; Franceschini et al. 1994; Sanders and Mirabel 1996). Mid- and far-infrared observations of the high-redshift universe are then needed for precise quantification of the dust-extinction effect: without a direct observation of the dust-reradiated part of the young stellar population emission, a reliable estimate of the start formation rate is difficult, or even impossible (e.g.  Poggianti \& Wu, 2000; Poggianti et al. 2001).

In the redshift interval from local up to $z\sim 1$, relatively deep IR observations have been obtained with early space infrared missions like IRAS (Soifer et al. 1988) and ISO (Genzel \& Cesarsky 2000).  Based on the fair sensitivity and moderate spectral coverage of the latter, the galaxy and AGN evolution in the redshift interval $0<z<1$ has been mapped (Franceschini et al. 2001; Chary \& Elbaz 2001; Rowan-Robinson 2001, 2009; Elbaz et al. 2002; Lagache et al. 2004; Pozzi et al. 2004), but there was insufficient sensitivity at higher redshifts for any significant constraints to be set there.

For a long time, both theoretical expectations (Franceschini et al. 1994, 1997; Blain \& Longair 1993; see also Kormendy \& Sanders 1992) and preliminary observational results by millimetric and sub-millimetric surveys with SCUBA (Smail et al. 2002; Hughes et al. 1998; Blain et al. 1999, 2002; Chapman et al. 2005) and IRAM (Greve et al. 2005; Bertoldi et al. 2007) have produced evidence that major events of star formation in the universe happened at high redshifts, $z>1$, and are likely responsible for the origin or assembly of the most massive galaxies today. Such events would be accompanied by major energy release, but, because of heavy dust extinction, would be detectable only through far-infrared observations. Sub-millimetric surveys indeed have identified sources that are extremely luminous at long wavelength, but almost undetectable in the optical (e.g. Chapman et al. 2005; Mobasher et al. 2006; Rodighiero et al. 2007).

A dramatic improvement in our knowledge of the infrared emissivity of source populations at redshifts $z>1$, corresponding to the the era of peak star formation and AGN activity, has recently been made possible with the deep sky exploration by the \textit{Spitzer Space Telescope} (SST, Werner et al. 2004). The Multiband Imaging Photometer (MIPS, Rieke et al. 2004) on SST combined sensitivity and spectral coverage, particularly with the $\lambda_{eff}=24\ \mu$m band, adequate for identifying and  characterizing substantial numbers of far-IR galaxies up to $z\simeq 3$ (Le Floch et al. 2005; Dole et al. 2006; Caputi et al. 2007; Lonsdale et al. 2004; Rowan-Robinson et al. 2005). The MIPS instrument also includes photometric imagers at $\lambda_{eff}=70$ and 160 $\mu$m (although the latter was limited by source confusion).
A further crucial contribution by the SST mission was the capability offered by the Infrared Spectrometer (IRS, Houck et al. 2004) to obtain high-quality far-IR spectra of high-redshift sources for the first time.

A historical difficulty in identifying and interpreting luminous high-redshift sources, i.e. the difficult association of the optical counterparts to the IR-mm sources due to the poor spatial resolution and the large IR error-boxes, is also largely overcome by observations with the \textit{Spitzer} Infrared Array Camera (IRAC, Fazio et al. 2004). IRAC deep multiwavelength imaging at 3 to 10 $\mu$m allows continuous band-merging and association of far-IR and optical/near-IR source catalogues in this critical spectral transition region (e.g. Egami et al., 2004).

The multiwavelength observations by \textit{Spitzer} have fully confirmed previous expectations of numerous populations of luminous galaxies and active nuclei at high redshifts. Sources at $z>1$ and up to $\sim3$ are routinely detected with the sensitive, high-spatial resolution 24 $\mu$m channel of Spitzer/MIPS (Lacey et al. 2004; Perez-Gonzales et al. 2005; Lacey et al. 2004; Caputi et al. 2006; Daddi et al. 2007). The MIPS imager benefits from a wide enough detector format to allow deep surveying with high mapping speed of several areas of relevant cosmological interest (Papovich et al., 2004; Lonsdale et al. 2004; Le Floch et al., 2004; Shupe et al. 2007; Sanders et al. 2007). 

MIPS surveys at longer wavelengths (70 and 160 $\mu$m) have produced more limited samples of distant sources (Frayer et al. 2006a,b; 2009), because of the much longer integration times required and the lower mapping speed at faint flux levels. These faint far-IR \textit{Spitzer} sources are also more difficult to identify, because of the poor \textit{Spitzer} resolution at such long wavelengths. A major step forward is now possible with the \textit{Herschel Space Observatory} (Pilbratt 2005; 2010).

The present paper is dedicated to a statistical analysis of a very extensive dataset of IR sources, aimed at deeper understanding of the cosmological significance of the long-wavelength emission in the complex process of galaxy formation and evolution.
A previous analysis by Franceschini et al. (2001, AF2001), based mostly on IRAS and ISO data, as well as on estimates of the cosmic IR background (CIRB) by COBE, has already achieved interesting results. These concerned in particular the evidence of a population of fast-evolving starburst galaxies between z=0 and 1 and the indication of a constant IR emissivity for galaxies at $z>1$. Results consistent with these were reported by Chary \& Elbaz (2001) and Elbaz et al. (2002). Another indication by AF2001 was in favour of a top-heavy stellar initial mass function (IMF) during major star-formation phases, to reconcile the high IR emissivity with local contraints on the galaxy stellar mass function.

All these had to be taken as preliminary hints, owing to the lack of any statistical data above $z\sim1$ available to AF2001.
A much more detailed and complete analysis is now possible based on the vast amount of new information available from \textit{Spitzer}, particularly at high redshifts. 
New essential data also include extensive sub-millimetric imaging surveys with BLAST (Pascale et al. 2008), and preliminary constraints by the \textit{Herschel} observatory. 
Thanks to the tight relationship of far-IR emissivity of galaxies with the instantaneous rate of star formation (Kennicut 1998; Poggianti et al. 2001; Elbaz et al. 2002), a model reproducing these observations of faint IR galaxies will provide us with a detailed description of the history of star formation in galaxies.
While other galaxy evolutionary models have recently been published (Lagache et al. 2004 and references therein), the current paper accounts for a lot of new, partly unpublished information, particularly on the redshift and luminosity distributions of faint Spitzer/MIPS sources, time-dependent luminosity functions, and background fluctuations, providing us with an accurate description of the high-redshift evolution.

The present paper also includes a first quantitative comparison of the history of the bolometric IR emissivity with the evolutionary stellar mass functions of galaxies (from deep \textit{Spitzer} IRAC near-IR surveys), entailing interesting constraints on the nature of the high-redshift luminous source population. We recall that, while the CIRB background intensity is a record of the past star-formation activity biased to low redshifts by the usual redshift factor $(1+z)^{-1}$ (e.g. Harwit 1999), the comoving mass density in galactic stars is a completely unbiased tracer that is much more sensitive to the high-redshift activity.

\begin{figure}[!ht]
\centering
\includegraphics[angle=0,width=0.5\textwidth,height=0.4\textwidth]{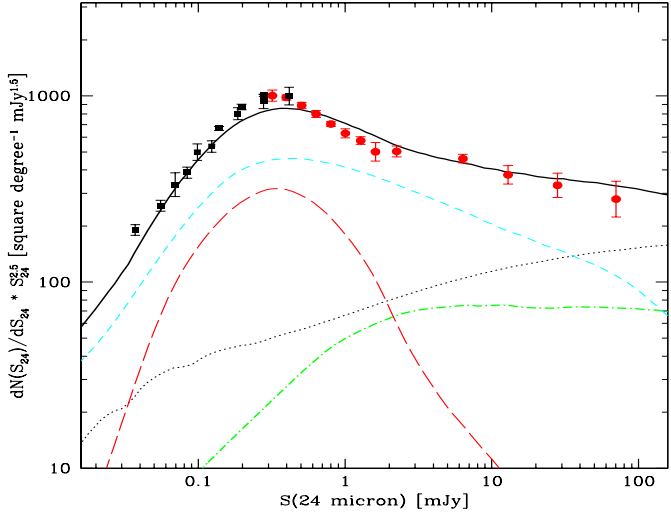}
\caption{Euclidean-normalized differential number counts of extragalactic sources at $24\mu$m compared with our model fit. 
The red circles are from the analysis of SWIRE survey data by Shupe et al. (2008), black squares from Papovich et al. (2004).
The contribution by type-I AGNs is shown as green dot-dashed line, moderate-luminosity starbursts (the LIRGs) make the cyan short-dash line (type-II AGNs and starbursts are included in the same population on the assumption that in both classes the IR spectrum is dominated by starburst emission).
The red long-dashed line corresponds to the population of high-luminosity sources dominating the IR emissivity at high redshifts.
The dotted line is the separate contribution of normal spirals, while the continuous line is the total model counts.
}
\label{c24}
\end{figure}

\begin{figure}[!ht]
\centering
\includegraphics[angle=0,width=0.5\textwidth,height=0.4\textwidth]{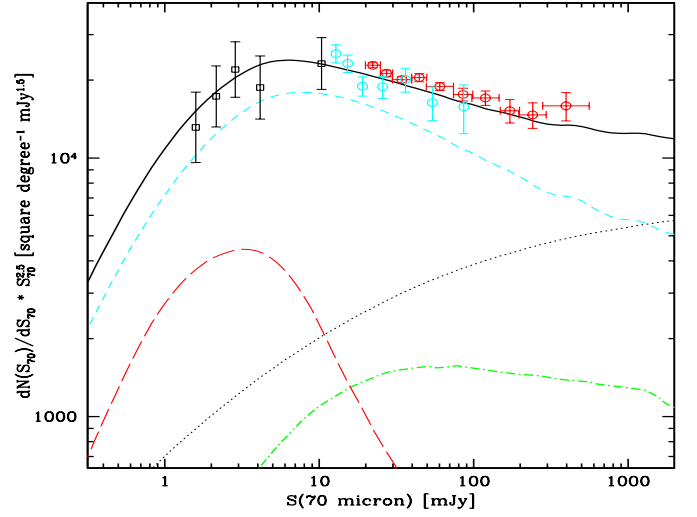}
\caption{Euclidean-normalized differential number counts of extragalactic sources at $70\mu$m compared with our model fit. 
Red open circles are from our analysis of the SWIRE survey data (Vaccari et al. in preparation, see text), open squares to a very deep survey by Frayer et al. (2006a), the cyan circles are the most recent assessment of the counts based on Spitzer/MIPS observations in the COSMOS area by Frayer et al. (2009).
The red filled datapoint at bright fluxes is from the IRAS 60 $\mu$m survey appropriately scaled to the current effective wavelength (Vaccari et al. 2009).
Moderate-luminosity starbursts (including type-II AGNs) make the cyan short-dash line.   
The red long-dashed line corresponds to the population of high-luminosity sources dominating the IR emissivity at high redshifts.
The dotted line is the separate contribution of normal spirals. At bright 70 $\mu$m flux densities type-I AGNs, shown as green dot-dashed line, provide an important contribution to the counts. 
The continuous line is the total model counts.
}
\label{c70}
\end{figure}

\begin{figure}[!ht]
\centering
\includegraphics[angle=0,width=0.5\textwidth,height=0.4\textwidth]{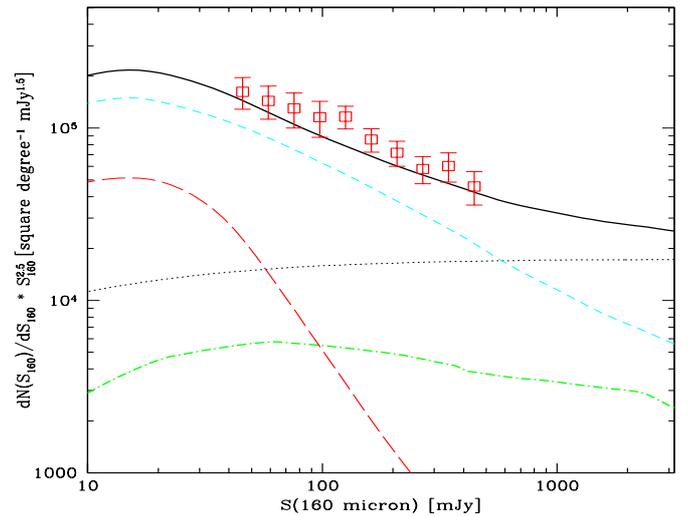}
\caption{Euclidean-normalized differential number counts of extragalactic sources at $160\mu$m compared with our model fit. Lines are as in Fig. \ref{c70}.  The red open square datapoints are from a re-analysis of the Spitzer/MIPS data by Bethermin et al. (2010).
}
\label{c160}
\end{figure}

The paper is structured as follows. 
Section 2 reports on the wide variety of observational data on distant IR sources making our reference dataset. These include some new estimates of source number counts from the SWIRE surveys at 70 and 160 $\mu$m and new evaluations of the redshift distributions of faint 24 $\mu$m sources. We also report new data here on local luminosity functions in Spitzer/MIPS bands, establishing a benchmark against which evolutionary models of cosmic sources are compared. 
In Sect. 3 we report on multi-wavelength models of galaxy and AGN evolution fitting the whole variety of existing data, including the local background spectral intensity, as well as data on distant and high redhift sources.  
In Sect. 4 we discuss our results and identify particular evolutionary patterns emerging from our analysis, including detailed comparison of the source comoving IR emissivity with the comoving stellar mass density and discuss implications from this comparison.
Our conclusions are summarized in Sect. 5. 
Two appendices are dedicated to discussing the contribution of discrete sources to the background intensity and cell-to-cell fluctuations and to elaborating some model predictions, with particular reference to the forthcoming \textit{Herschel} observations and source \textit{confusion} limitations that will be encountered.

All quantities are computed in this paper assuming a geometry for the Universe with $H_0=70$ km s$^{-1}$ Mpc$^{-1}$, $\Omega_m=0.3$, $\Omega_\Lambda=0.7$.
We indicate with the symbols $S_{24}$ and $L_{24}$ the flux density in Jy at 24 $\mu$m and the monochromatic luminosity $\nu L_\nu$ at 24 $\mu$m in erg/s normalized to the solar value $L_\odot$ (and similarly for other wavelengths).


\section{The dataset}

Our analysis reported in the present paper has benefited from the availability of highly reliable deep extragalactic surveys in a variety of IR bands, thanks mostly to the sensitivity and photometric accuracy of the \textit{Spitzer Space Telescope}'s MIPS instrument (Rieke et al. 2004). These properties of the far-IR camera have allowed not only for probing to very deep flux limits ($S_{24} \simeq 20\ \mu Jy$) in the 24 $\mu$m channel, but also complementing these ultradeep maps with large-area surveys covering several tens of square degrees, still to appreciable depths. 

The main \textit{Spitzer}'s limitation rests in the poor spatial resolution offered by the diffraction-limited photometric imagers in the critical far-IR (70 to 160 $\mu$m) spectral region, which implies a strong confusion noise, particularly at 160 $\mu$m. This limit is currently being overcome by the ongoing operation of the \textit{Herschel Space Observatory}.
We are also recovering important constraints on the high-redshift galaxy emissivity at long wavelengths from deep millimetric and sub-mm observations with large ground-based observatories (JCMT, IRAM, CSO among others).

\subsection{The IR multi-wavelength source number counts}

\subsubsection{The Spitzer MIPS 24 $\mu$m counts}

The \textit{Spitzer} Wide-Area Infrared Extragalactic (SWIRE) survey (Lonsdale et al. 2003, 2004), the largest of the \textit{Spitzer} extragalactic Legacy Science programmes, mapped 49 square degrees split into 6 widely separated fields, at wavelengths from 3.6 to 160 microns using both IRAC and MIPS. 
The MIPS 24 $\mu$m counts were derived by Shupe et al. (2008) from all six SWIRE's areas to minimize and averaged out sample variance effects. 
Indeed, at fluxes brighter than $S_{24} \simeq 2\ mJy$, large field-to-field variations, by up to a factor of two, are found in the counts, whereas they showed differences on the order of only 10\% in the sub-mJy region.
The Euclidean-normalized differential counts reported by Shupe et al. (2008) averaged over the six independent SWIRE fields are reported in Figure \ref{c24}.

\begin{figure*}
\begin{minipage}{0.24\textheight}
\resizebox{6.2cm}{!}{
\includegraphics{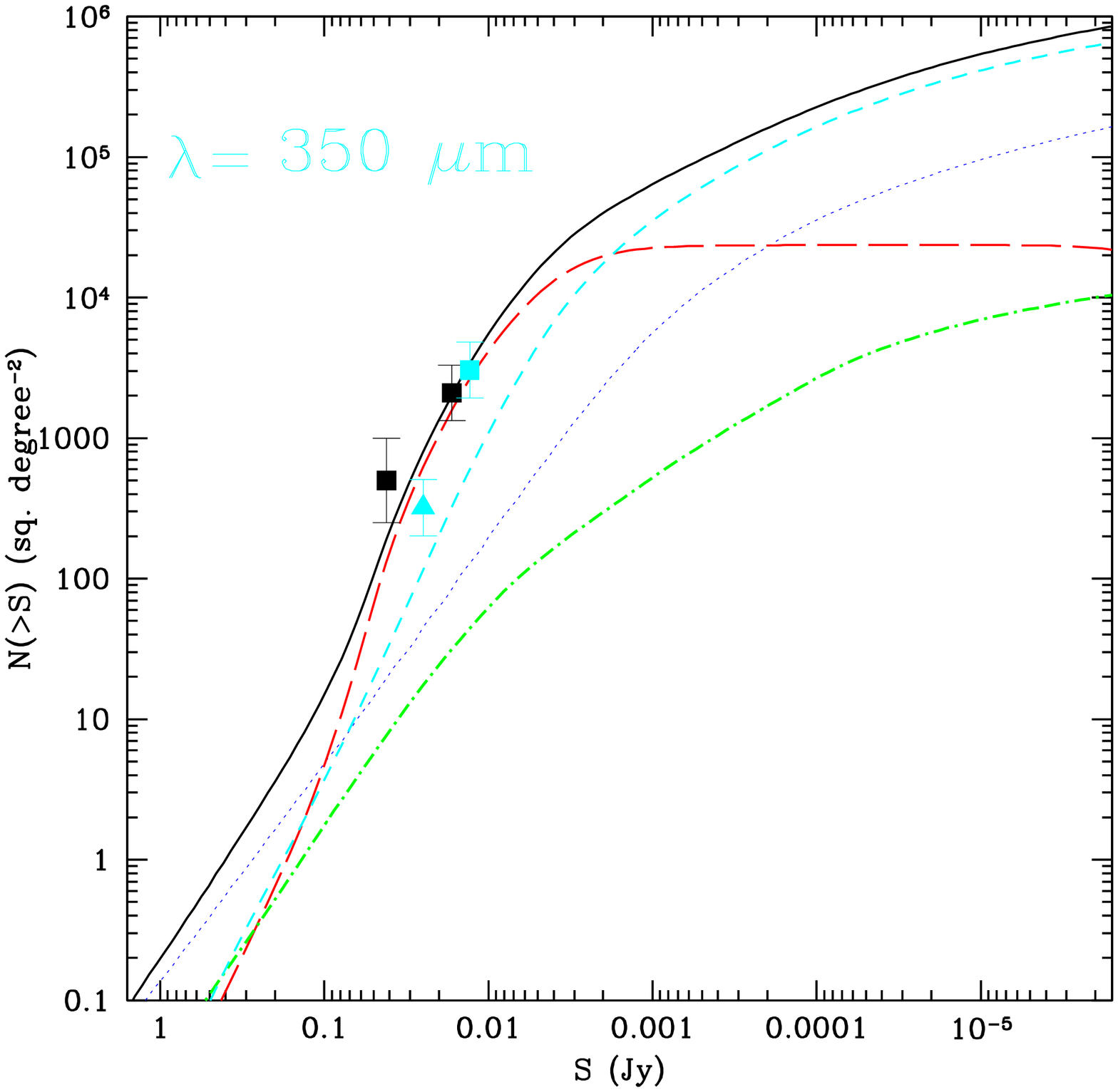}} 
\end{minipage}
\begin{minipage}{0.24\textheight}
\resizebox{6.2cm}{!}{
\includegraphics{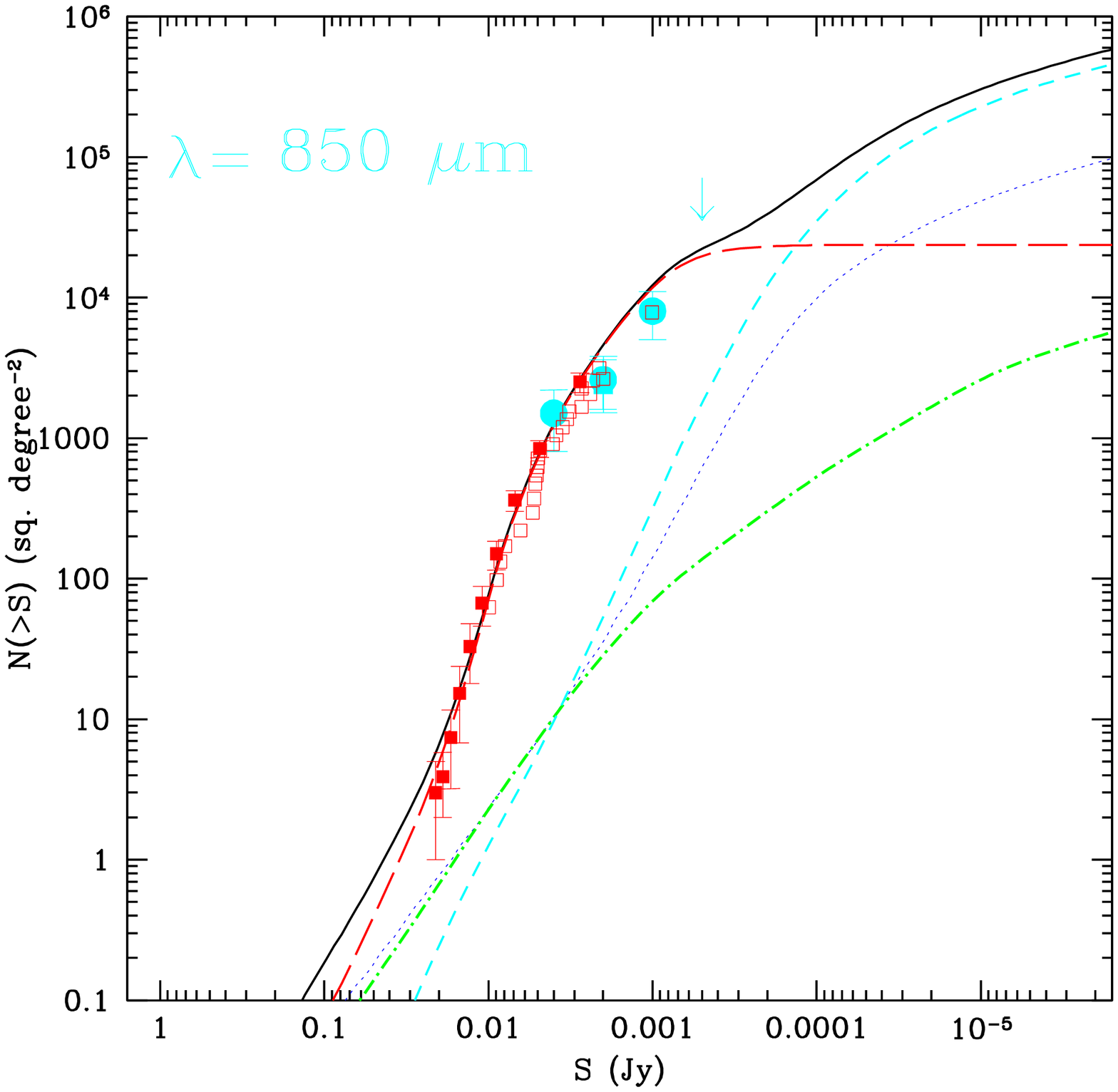}}
\end{minipage}
\begin{minipage}{0.24\textheight}
\resizebox{6.2cm}{!}{
\includegraphics{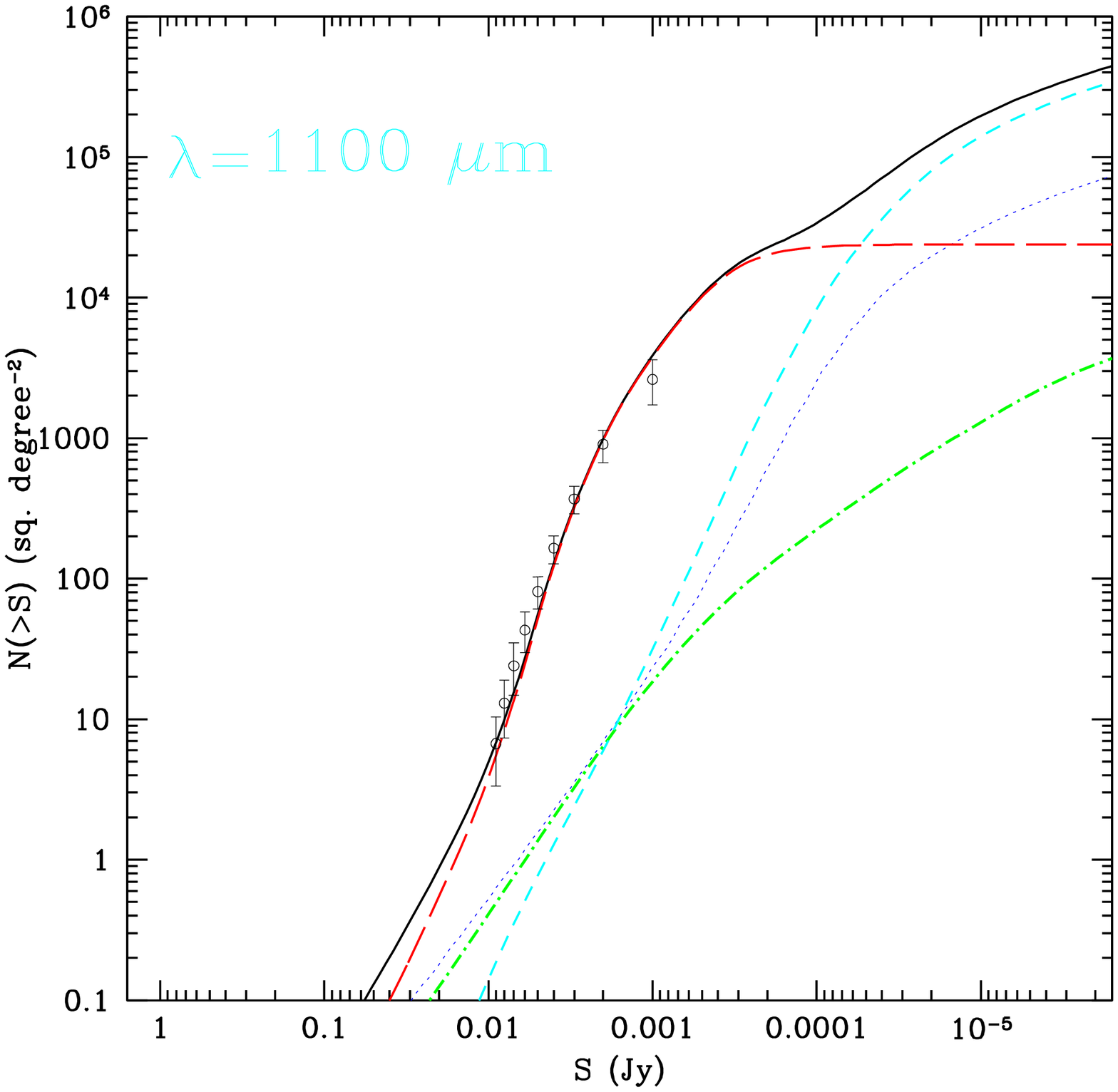}}
\end{minipage}
\caption{\textit{Left: } a collection of observed number counts at 350 $\mu$m by Khan et al. (2007) compared with our best-fit model prediction.
\textit{Center:} the observed number counts at 850 $\mu$m by Blain et al. (1999) and Barger, Cowie and Sanders (1999, open squares). Filled squares are a recent assessment by Coppin et al. (2006). 
\textit{Right: } number counts at 1100 $\mu$m by Austermann et al. (2009) from an Aztec survey at JCMT in the COSMOS area.
Line types as in Fig. \ref{c70}.
 } 
\label{mm}
\end{figure*}

Deeper extragalatic number counts, down to flux limits of $S_{24} \simeq 20\ \mu Jy$ with the MIPS 24 $\mu$m channel, have been obtained in the GOODS, GTO and COSMOS surveys by Papovich et al. (2004), Chary et al. (2004), Marleau et al. (2004), Rodighiero et al. (2006), Sanders et al. (2007).
A selection of these deeper counts is reported in Fig. \ref{c24}. It is interesting to note the perfect agreement among the results of different teams, sky areas, and data reduction procedures required corrections to the total fluxes of only few \% altogether, showing the high reliability of MIPS at these wavelengths.

Altogether, the number counts show a steady super-Euclidean slope as a function of flux density all the way down to the SWIRE completeness limit of 0.4 mJy, with a hint of convergence below it. At fainter fluxes the counts from the deeper surveys show a very fast convergence down to the MIPS confusion limit, estimated to fall around a few tens of $\mu Jy$. A similar \textit{bumpy} behaviour, though not as much extreme, was observed by Elbaz et al. (1999) in the deep ISO counts at 15 $\mu$m and was interpreted as a signature of very fast evolution in the galaxy emissivity from z=0 to 1 (see Sect. 3).

\subsubsection{The \textit{Spitzer} MIPS counts at 70 and 160 $\mu$m }
\label{MIPS}

MIPS observations at longer wavelengths required much more extensive integrations to achieve comparable depths. 
Deep surveys have then only been possible on small sky areas, the bulk of the observations in the SWIRE and FLS fields having been performed to relatively shallow depths. 
The deepest observations at 70 and 160 $\mu$m were also limited by source confusion. 

We used the SWIRE database to estimate the extragalactic number counts at 70 $\mu$m. While a detailed account of this work will be reported in a separate paper (Vaccari et al., 2009, and paper in preparation), we summarize here the main facts. Source catalogues at 24 and 70 $\mu$m were extracted by us in the six areas observed by the SWIRE survey over a total of 49.5 square degrees, where good complementary data exist to perform an accurate subtraction of stars. As a consequence of the quite different sensitivities, the vast majority of the 70 $\mu$m sources had counterparts at 24  $\mu$m. In turn, essentially all the 24 $\mu$m objects are identified in the short wavelength IRAC maps. This has prompted us to obtain highly reliable catalogues at these wavelengths. In all, Vaccari et al. identified about 10000  extragalactic sources at 70 $\mu$m brighter than the 5$\sigma$ limit, respectively.
The survey completeness and source reliability were extensively tested by inserting simulated sources into the real maps. The corresponding Euclidean-normalized differential counts from the SWIRE survey are reported in Figs. \ref{c70}. These data compare well with those derived by Frayer et al. (2006b) from the FLS survey. 

Deeper number counts at 70 $\mu$m, based on a very long MIPS integration in the central 10'x10' area of GOODS-N and a shallower GTO observation over 0.6 sq.deg. in GOODS-N, are reported in Frayer et al. (2006a) and shown in Fig. \ref{c70}. A survey at 70 and 160 $\mu$m on a much wider COSMOS 2 square degree area is reported by Frayer et al. (2009). These observations have prompted them to accuratele evaluate the extragalactic source confusion noise at 70 $\mu$m, which is estimated to be around $0.35\pm 0.15$ mJy, and correspondingly estimated the counts down to a $4\sigma$ flux limit of 1.2 mJy.
 
We report in Fig. \ref{c160} the results of a systematic re-analysis of Spitzer/MIPS number counts data by Bethermin et al. (2010).
At 160 $\mu$m, the \textit{Spitzer} 1$\sigma$ confusion noise sets around $\sigma_{confusion} \approx 9.4 \pm 3.3$ mJy according to Frayer et al. (2009) and Dole et al. (2006, see also Sect. \ref{discussion} below). The flux limit of $\sim 40$ mJy in the number count data of Fig. \ref{c160} is then close to that set by confusion. This issue of the source confusion noise for the Spitzer/MIPS observations will be  discussed later in Sect. \ref{pdd} when talking of background fluctuations.

Interesting features can be found in Figs. \ref{c70} and \ref{c160}. The 70 $\mu$m ultradeep counts reveal  fast convergence at faint fluxes, although with limited statistics, similar to those observed at 24 $\mu$m (Fig. \ref {c24}) and 15 $\mu$m (AF2001). This implies that the majority of the extragalactic background light (EBL) has already been resolved into sources by Spitzer at this wavelength and that
a major part of the radiant energy produced by sources has been identified. In particular, about 60\% of the total estimated EBL at 70 $\mu$m has been resolved into sources by Frayer et al. (2006a) down to the confusion limit, and as much as 75\% of it at 24 $\mu$m.

Also the slopes of the differential counts in `the super-Euclidean' part at moderately bright fluxes up to the peak bring interesting information. These counts are flatter at 70 $\mu$m and steeper at both 24 and 160 $\mu$m, as a consequence of the very different K-corrections for the average galaxy spectra at the respective wavelengths: the K-correction is minimal at 70 $\mu$m owing to the very steep spectra towards short wavelengths. This also entails different average source redshifts for the flux-limited samples at the two wavelengths.

\subsubsection{The CSO, SCUBA, and IRAM counts at 350, 850, and 1200 $\mu$m }

Since the early observations with SCUBA (Ivison et al. 1998), later confirmed with IRAM and CSO, deep surveys with millimetric and sub-millimetric observatories have contributed some fundamental insight into the high-redshift universe, by showing the existence of extremely luminous source populations that were essentially undetected in the optical. 
The identification of the optical counterparts of these objects has turned out to be a painful process, because of the large positional error-box of mm telescopes and the source faintness (Blain et al. 2002). Eventually, such a process required a simultaneous identification in the radio (Chapman et al. 2005), or the high-resolution imaging with mm interferometers (e.g. IRAM-PdB, Tacconi et al. 2006), which are available for few sources. The situation for astronomy at millimetric wavelengths will not change significantly until the advent of ALMA.

The limited information about statistical properties, like the high-z luminosity functions or redshift distributions of the mm sources, led us to mostly use the information in the number counts to constrain our model at long wavelengths (hence high redshifts, due to large K-correction).
Figure \ref{mm} compares the results of our model with the integral counts at 350, 850, and 1100 $\mu$m.
A collection of the 350 $\mu$m counts is reported by Kahn et al. (2007) based on deep CSO observations.

Number counts at 850 $\mu$m with SCUBA on JCMT (central panel of Fig. \ref{mm}) have been reported by Blain et al. (1999) and Barger, Cowie, and Sanders (1999).  The most statistically accurate recent evaluation has been published by Coppin et al. (2006) based on the largest, most complete and unbiased extragalactic submillimetre survey, the SCUBA HAlf Degree Extragalactic Survey (SHADES). The actual survey covered 720 arcmin$^2$ to 2 mJy $1\sigma$, detecting approximately a hundred sources. The differential number counts by Coppin et al. are reported in Fig. \ref{dz850} below.

Counts at millimetric wavelengths (1100-1200 $\mu$m) from IRAM MAMBO and BOLOCAM have been published by various authors.  The right panel of Fig. \ref{mm} reports a recent assessment of the number counts at 1100 $\mu$m by Austermann et al. (2009) from a survey with Aztec at JCMT in the COSMOS area.

\subsubsection{The BLAST sub-millimetric observations }
\label{bla}

The balloon-borne BLAST mission (Pascale et al. 2008) has recently performed an exploratory survey of a region in CDFS using a similar instrumental setup and the same waveband filters that are used by the Herschel Space Observatory at these wavelengths. Devlin et al. (2009) and Patanchon et al. (2009) estimate source number counts based on a P(D) analysis from deep survey data in CDFS (0.8 sq.deg.) combined with shallower data over 8.7 sq.deg. A comparison of the BLAST estimated counts at 250, 350, and 500 $\mu$m with predictions of our reference model (Sect. \ref{model} below) is reported in Fig. \ref{blast}.

The BLAST survey data provide us with the first extensive imaging survey of the sky at sub-millimetric wavelengths, hence imposing an essential constraint on the source emissivities at wavelengths close to the peak of the CIRB extragalactic background. 
The uncertainties related to source confusion led us to consider the faintest datapoints at 10 to 20 mJy at the three BLAST wavelengths as upper limits on the real extragalactic counts.


\begin{figure}
\includegraphics[angle=0,width=0.5\textwidth,height=0.4\textwidth]{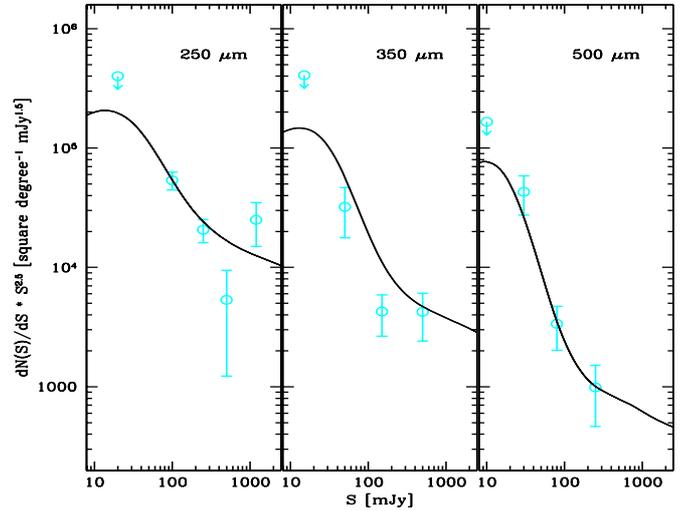}
\caption{Data on the Euclidean-normalized differential number counts of sources at three sub-millimetric wavelengths obtained by Patanchon et al. (2009, see also Devlin et al. 2009) from observations with BLAST in the CDFS area. The two BLAST 250 $\mu$m datapoints at the faintest fluxes have been combined in our plot. The black continuous line is the total predicted counts by our reference model.
} 
\label{blast}
\end{figure}

\subsubsection{Preliminary constraints from Herschel observations }
\label{H}

After its successful launch in May 2009, instrument validation and Performance Verification, the Herschel Space Observatory carried out Science Demonstration observations during the second semester of 2009 and has subsequently started the routine observation phase. 
Herschel carries on-board two sensitive photometric imaging cameras, PACS and SPIRE, covering a wide wavelength interval between 70 and 500 $\mu$m, where the bulk of the CIRB energy resides.
While a first systematic account of cosmological observations with Herschel will be reported in a dedicated issue of this Journal, some constraints can already be inferred from the Science Demonstration Phase Workshop in December 2009 (see \texttt{http://herschel.esac.esa.int/SDP$\_$IR$\_$wkshop.shtml}).
At the present stage we have confined ourselves to checking that the data appearing in those reports (particularly those by S. Oliver and D.Lutz) do not conflict with our model results.  At the same time, these data appear to agree closely with the BLAST results at the corresponding wavelengths.

\begin{figure}[!ht]
\centering
\includegraphics[angle=0,width=0.5\textwidth,height=0.4\textwidth]{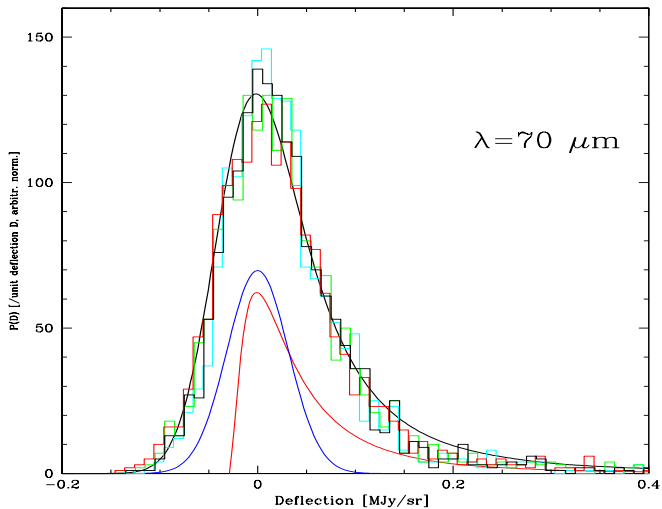}
\caption{Histogram of the distribution of background intensity in MJy/sr measured inside circular apertures of 12, 14, and 16 arcsec radius from the FIDEL 70 $\mu$m map (Frayer et al. 2009, green, cyan, red histograms respectively). The observed histograms of deflections per unit sky area $D_a$ are compared with a Gaussian distribution representing the instrumental noise, having an rms of 0.62 mJy/beam (blue line), and with the probability distribution of deflections $D_a$  due to sources, which are modelled according to our best-fit source-count model (red line). The black line is a convolution of the two and provides a good fit to the observational distribution.
}
\label{pdd70}
\end{figure}

\begin{figure}[!ht]
\centering
\includegraphics[angle=0,width=0.5\textwidth,height=0.4\textwidth]{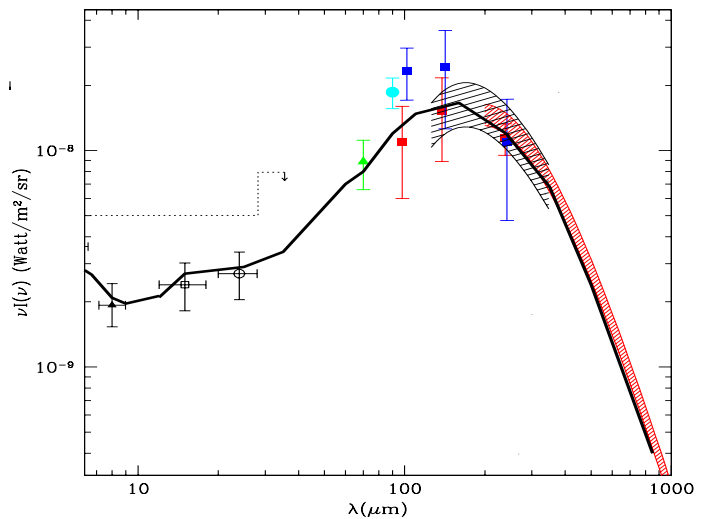}
\caption{Data on the cosmic infrared background (CIRB; e.g. Hauser et al. 1998) spectral intensity compared with the best-fit model predictions.
The three lower datapoints in the far-IR are from a re-analysis of the DIRBE data from the all-sky COBE maps by Lagache et al. (1999), the shaded areas from Fixsen et al. (1998, black shade) and Lagache et al. (2004, red shade). 
The two mid-IR datapoints at 15 and 24 $\mu$m are the resolved fraction of the CIRB by Elbaz et al. (2002) and Papovich et al. (2004). 
The triangle datapoint at 8 $\mu$m comes from the integration of the IRAC galaxy counts by Fazio et al. (2004) and Franceschini et al. (2008), while the green triangle at 70 $\mu$m is from the ultra-deep Spitzer survey by Frayer et al. (2006) (recalibrated by a scaling factor of 1.2 as indicated by Frayer et al. 2009).
The dotted histograms are upper limits on the CIRB intensity set by TeV cosmic opacity measurements (Stanev and Franceschini 1998).
The thick line is the predicted CIRB spectrum from the multi-wavelength reference model for galaxy evolution discussed in the present paper. 
}
\label{bkg}
\end{figure}

\begin{figure}[!ht]
\centering
\includegraphics[angle=0,width=0.5\textwidth,height=0.4\textwidth]{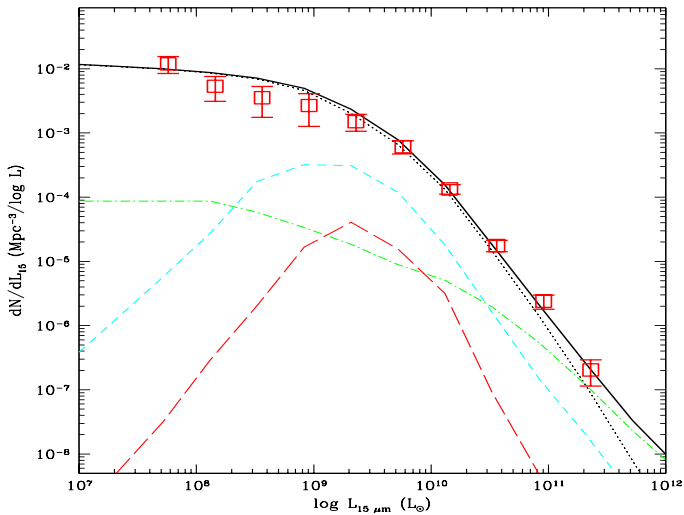}
\caption{Galaxy LLF's at $15\ \mu m$ from Fang et al. (1998) and adapted from Xu et al. (1998, red open squares) in the low-luminosity regime. Lines correspond to our best-fit model described in Sect. 4.
The contribution to the $15\ \mu m$ LLF by type-I AGNs is shown as green long-dashed line, type-II AGNs plus moderate-luminosity starbursts make the green short-dashed line (type-II AGNs and starbursts are included in the same population on the assumption that in both classes the IR spectrum is dominated by starburst emission).
The red line corresponds to the local LF of a population of high-luminosity starbursts dominating the source emissivity at high redshifts.
The dotted line is the separate contribution of normal spirals, while the continuous line is the total LLF.
}
\label{llf12}
\end{figure}

\subsection{The background fluctuation distribution $P(D)$ as a further constraint on the faint source statistics}
\label{pdd}

The analysis of the fluctuations in the background intensity provides other useful constraints on the faint source distribution in circumstances where the map's spatial resolution is poor, such that relatively few sources are directly detectable above the limits imposed by the instrumental and source confusion noises. With the term source confusion we refer to the \textit{photometric} confusion in the terminology of Dole, Lagache, \& Puget (2003).
The distribution of background fluctuations in sky cells, often called distribution $P(D)$ of the total flux (or "deflection" D) in the cell, includes information on the source number counts down to fluxes corresponding to one source per beam, well below the nominal confusion limit of at least 10-20 beams per source usually adopted (Scheuer 1974).
We defer the reader to a detailed account in Appendix A of the computational procedures to calculate the predicted $P(D)$ based on a model for the differential number counts and under the assumption of a random distribution of sources on the sky.

An interesting case is offered by the Spitzer/MIPS observations obtained by the FIDEL project (Frayer et al. 2009; Magnelli et al. 2009). The $P(D)$ analysis offers an important consistency check of the counts in Fig. \ref{c70} at fluxes fainter than 10 mJy, where the direct counts statistics is poor, and limited to the sources detected in GOODS-N by Frayer et al. (2006a).

Figure \ref{pdd70} shows a comparison of our model distribution of fluctuations D  with that of background deflections measured inside circular apertures of different radii (from 12 to 18 arcsec) from the COSMOS 70 $\mu$m map (Frayer et al. 2009). 
The observed distribution of deflections $D$ is considered here to stem from the convolution of the instrumental noise (assumed Gaussian) and the source confusion distribution.
The lines corresponding to the Gaussian noise distribution (rms of 0.62 mJy/beam, blue curve) and the probability distribution of deflections D due to source confusion are shown. The convolution of the two provides a good fit to the observational distribution.  The instrumental noise of 0.62 mJy/beam corresponds to 0.033 MJy/sr for our beams of 450 square arcsec, and the Gaussian instrumental noise is essentially set by the "negative" part of the $P(D)$ distribution (the side corresponding to the negative values of D) in Fig. \ref{pdd70}.  
Our estimate of the source confusion noise of $\sigma_{confusion}\sim 0.63$ mJy is larger than found by Frayer et al. (2009), 0.35$\pm$0.15 mJy, although only at the 2$\sigma$ confidence level. We do not have a clear explanation for this difference, except perhaps in the Frayer et al. procedure to perform a simple quadratic subtraction of the Gaussian instrumental noise from the total observed P(D), instead of a full deconvolution as done here, more appropriate for skewed distributions.

This analysis of the fluctuation distribution is important to confirm the fast convergence of the counts fainter than a few mJy in the 70 $\mu$m counts: a shallower convergence of the counts (see Fig. \ref{c70}) would imply excess signal in the P(D) distribution at $\mu >0.05\ MJy/sr$ compared to what is observed in Fig. \ref{pdd70}.  The P(D) data reported in the figure impose quite a fast convergence of these counts fainter than 10 mJy.

\subsection{The CIRB background intensity}
\label{bkgi}

Studies of galaxy evolution in the IR benefit by  including data on the integrated emission of all cosmic sources, in the form of a diffuse background radiation (the CIRB) detected in COBE all-sky maps between 100 and 1000 $\mu$m (Puget et al. 1996; Hauser et al. 1998; Lagache et al. 2004). We defer to these papers and to Franceschini et al. (2008) for reviews on the subject. 

A summary of the available data on the CIRB intensity $I(\nu)$, compared with our best-fit model predictions, is given in Fig. \ref{bkg}. Those in the far-IR and sub-mm are direct detections of the CIRB based on COBE maps, once the zodiacal and galactic foregrounds are subtracted. The mid-IR datapoints instead are based on simple integrations of the number counts in the respective wavebands published by Elbaz et al. (2002), Papovich et al. (2004), and Fazio et al. (2004).

These sub-mm CIRB detections are relevant for constraining the high-redshift evolution of the galaxy IR emissivity. Two effects balance together here: that of the K-correction, which is particularly strong because of the very steep sub-mm spectra of all galaxy types ($S[\nu]\propto \nu^{3.5}$), increases the source contribution at the high-z. On the other hand, for a given source comoving emissivity $\rho_{IR}(z)$ at redshift $z$, its contribution to the background intensity decreases proportionally to redshift as 
\[
\nu I(\nu) = c \rho_{IR}/4\pi(1+z) .
\] 
%

\begin{figure}[!ht]
\centering
\includegraphics[angle=0,width=0.5\textwidth,height=0.4\textwidth]{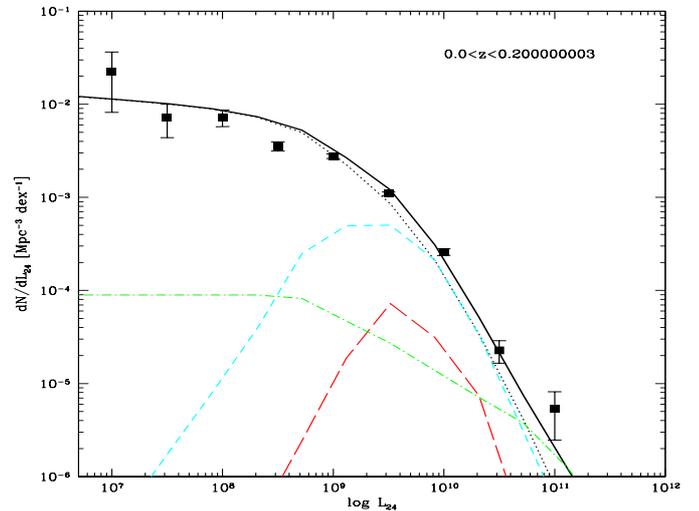}
\caption{Galaxy LLF's at $24\ \mu m$ from our analysis of the combined SWIRE-SDSS samples in the Lockman Hole (Vaccari et al. 2009). The contribution to the $24\ \mu m$ LLF by type-I AGNs is shown as a green long-dashed line, type-II AGNs plus starbursts make the green short-dash line. Type-II AGNs and starbursts are included in the same population on the assumption that the IR spectrum is dominated in both classes  
by starburst emission.
The dotted line is the separate contribution of normal spirals, while the continuous line
is the total LLF. 
}
\label{llf24}
\end{figure}

\subsection{The local boundary: luminosity functions}
\label{llf}

The local luminosity functions (LLF's) for galaxies selected at different IR wavelengths set a fundamental constraint on the population's cosmological evolution. Various independent estimates of the galaxy luminosity functions at different IR wavelengths have been reported. In the mid-IR, LLF's have been published by Rush et al. (1993), Xu et al. (1998) and Fang et al. (1998), all based on the 12 $\mu m$ IRAS survey from which statistically complete and numerous all-sky samples have been obtained. However, as discussed in AF2001, these were affected by various uncertainties, including the IRAS photometric errors and the effect of local inhomogeneities, particularly those from the the local supercluster, implying an improper flux normalization and too steep a faint-end slope (e.g. inconsistent with the IRAS 60 $\mu m$ LLF by Saunders et al. 1990). 
These various effects have been discussed by Fang et al. (1998), Xu et al. (1998), and AF2001: a corrected and re-calibrated 15 $\mu m$ luminosity function based on these analyses is reported in Fig. \ref{llf12} (see also AF2001).  
This re-estimate shows in particular a shallower low-luminosity slope that turned out to be consistent with the 60$\mu$m LLF by Saunders et al. (1990).

%
%

We have obtained new improved estimates of the IR multi-wavelength luminosity functions by cross-correlating the SWIRE survey (Lonsdale et al. 2003, 2004) in a large (10 sq.deg.) area in the Lockman Hole with the Sloan Digital Sky Survey (SDSS, Stoughton et al. 2002) spectroscopic and photometric redshifts. The combination of the high levels of photometric accuracy, completeness, and reliability of the SWIRE source catalogues and the completeness of the SDSS redshift information provided us with a very reliable source list for luminosity function studies. Also, the large area covered by the combined surveys in Lockman appears free of conspicuous local galaxy condensations and offers an unbiassed sample of the local universe. While all details about this analysis will be reported in Vaccari et al. (in progress), we summarize here the main facts.

\begin{figure}[!ht]
\centering
\includegraphics[angle=0,width=0.5\textwidth,height=0.4\textwidth]{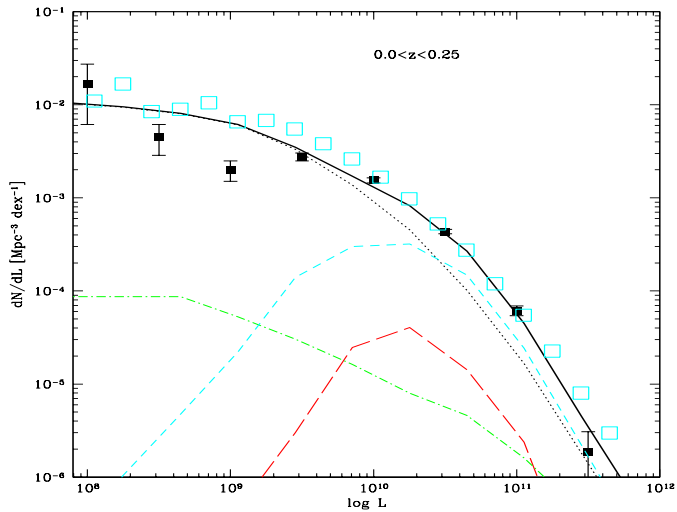}
\caption{
Galaxy LLF's at $70\ \mu m$ from our analysis of the combined SWIRE-SDSS samples in the Lockman Hole. The black filled square datapoints are as in Fig. \ref{llf24}, cyan open squares are the 60 $\mu$m local LF by Saunders et al. (1990) scaled to 70 $\mu$m with a typical galaxy SED. Lines as in Fig. \ref{llf12}.
}
\label{llf70}
\end{figure}

\begin{figure}[!ht]
\centering
\includegraphics[angle=0,width=0.5\textwidth,height=0.4\textwidth]{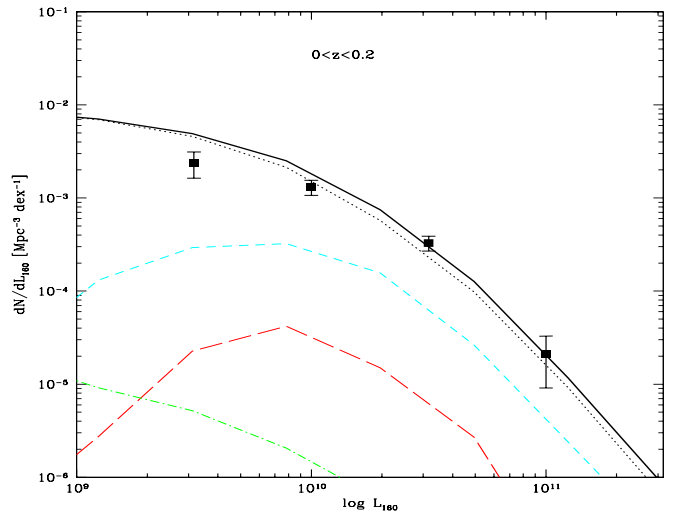}
\caption{Galaxy LLF's at $160\ \mu m$ from our analysis of the combined SWIRE-SDSS samples in the Lockman Hole.  Lines as in Fig. \ref{llf12}.
}
\label{llf160}
\end{figure}

\begin{figure*}
\begin{minipage}{0.38\textheight}
\resizebox{8.2cm}{!}{
\includegraphics{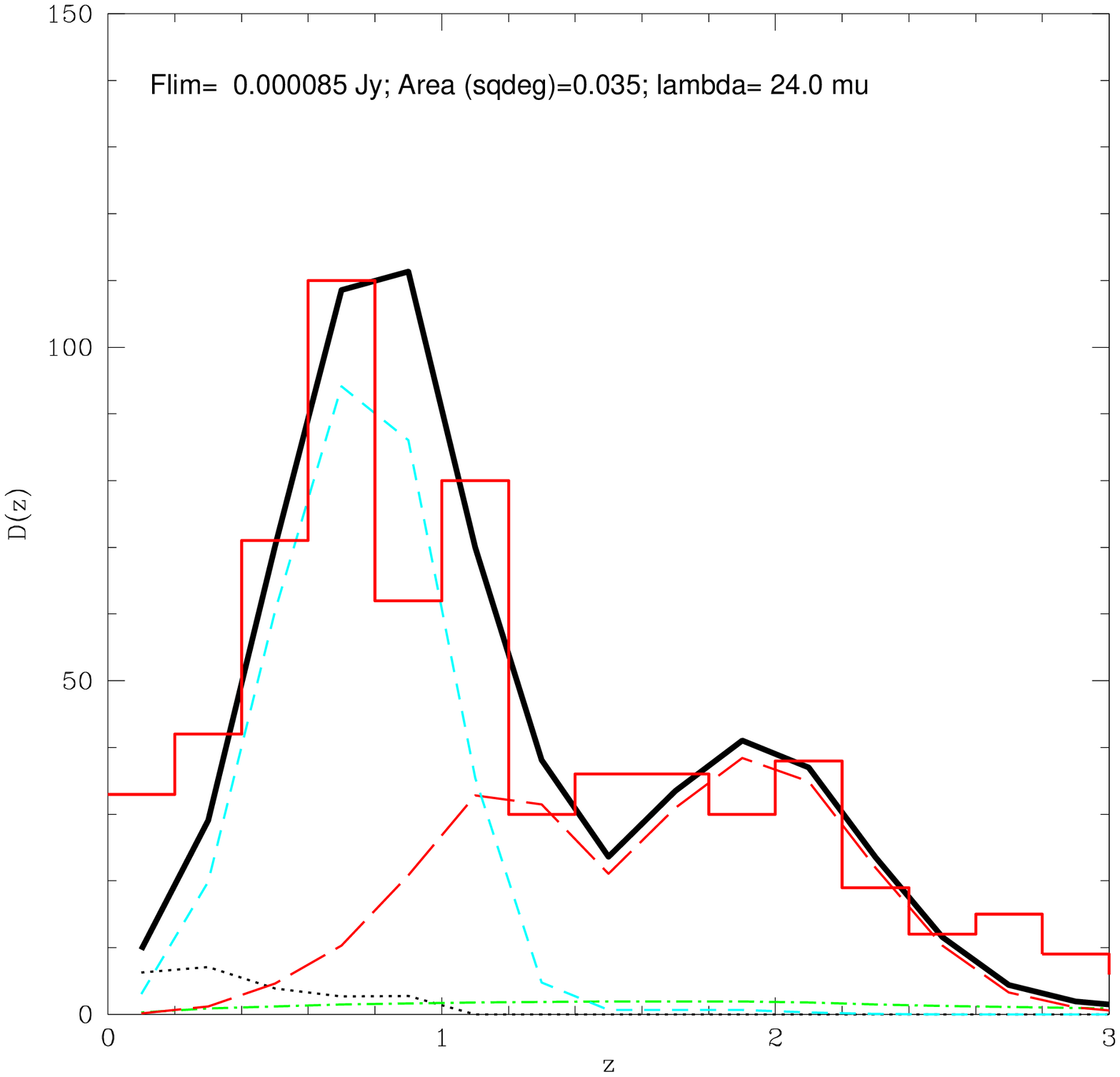}} 
\end{minipage}
\begin{minipage}{0.38\textheight}
\resizebox{8.2cm}{!}{
\includegraphics{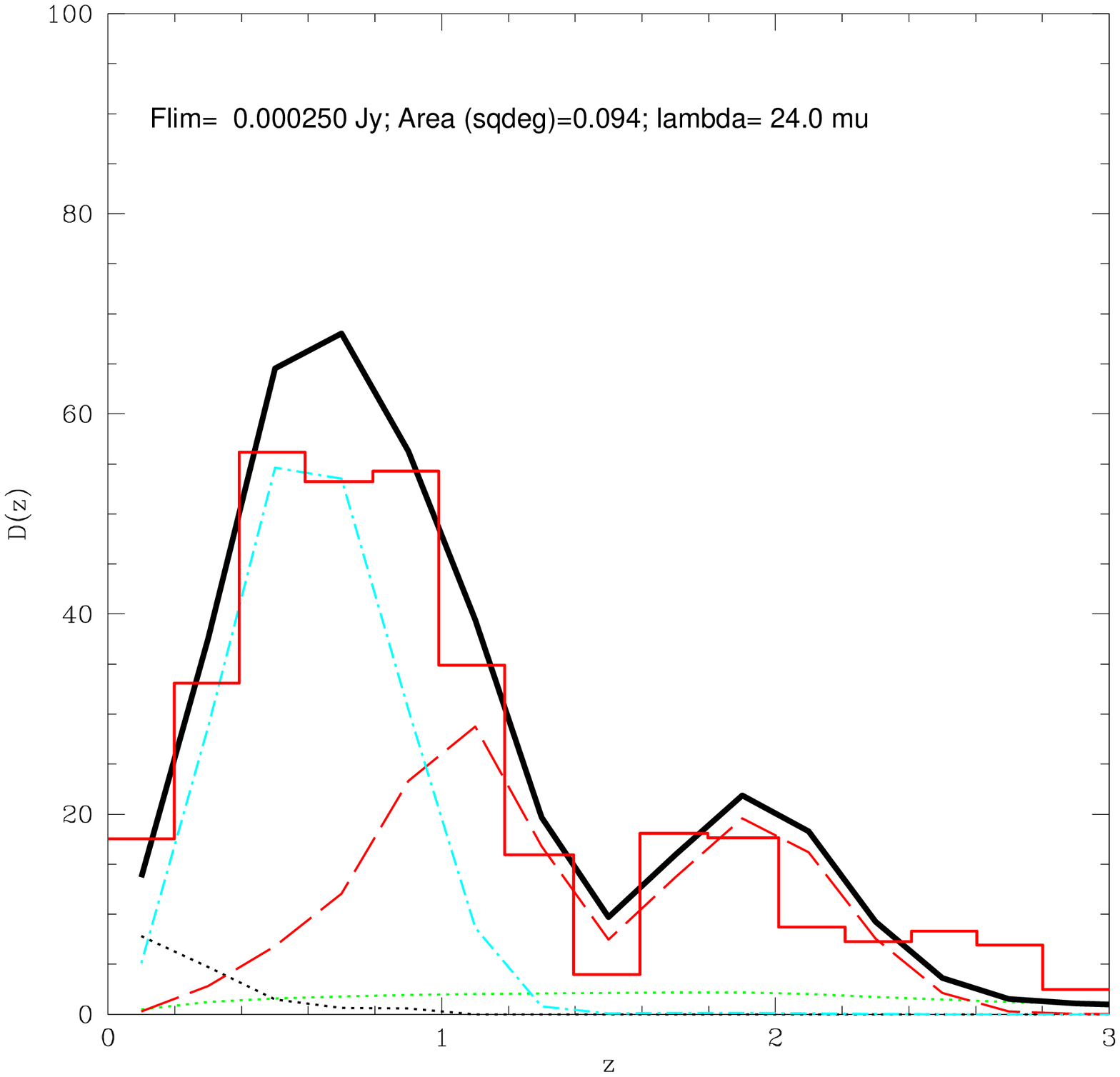}}
\end{minipage}
\caption{\textit{Left: } The observed redshift distributions of 24 $\mu$m sources with $S_{24}> 80\ \mu Jy$ in the GOODS-S field (red histogram). 
\textit{Right: } same as in the left panel but for a brighter flux limit $S_{24}> 250\ \mu Jy$ in the total area of 348 sq.arcmin covered by GOODS-S and GOODS-N.
 } 
\label{dz24}
\end{figure*}

\begin{figure*}
\begin{minipage}{0.38\textheight}
\resizebox{8.2cm}{!}{
\includegraphics{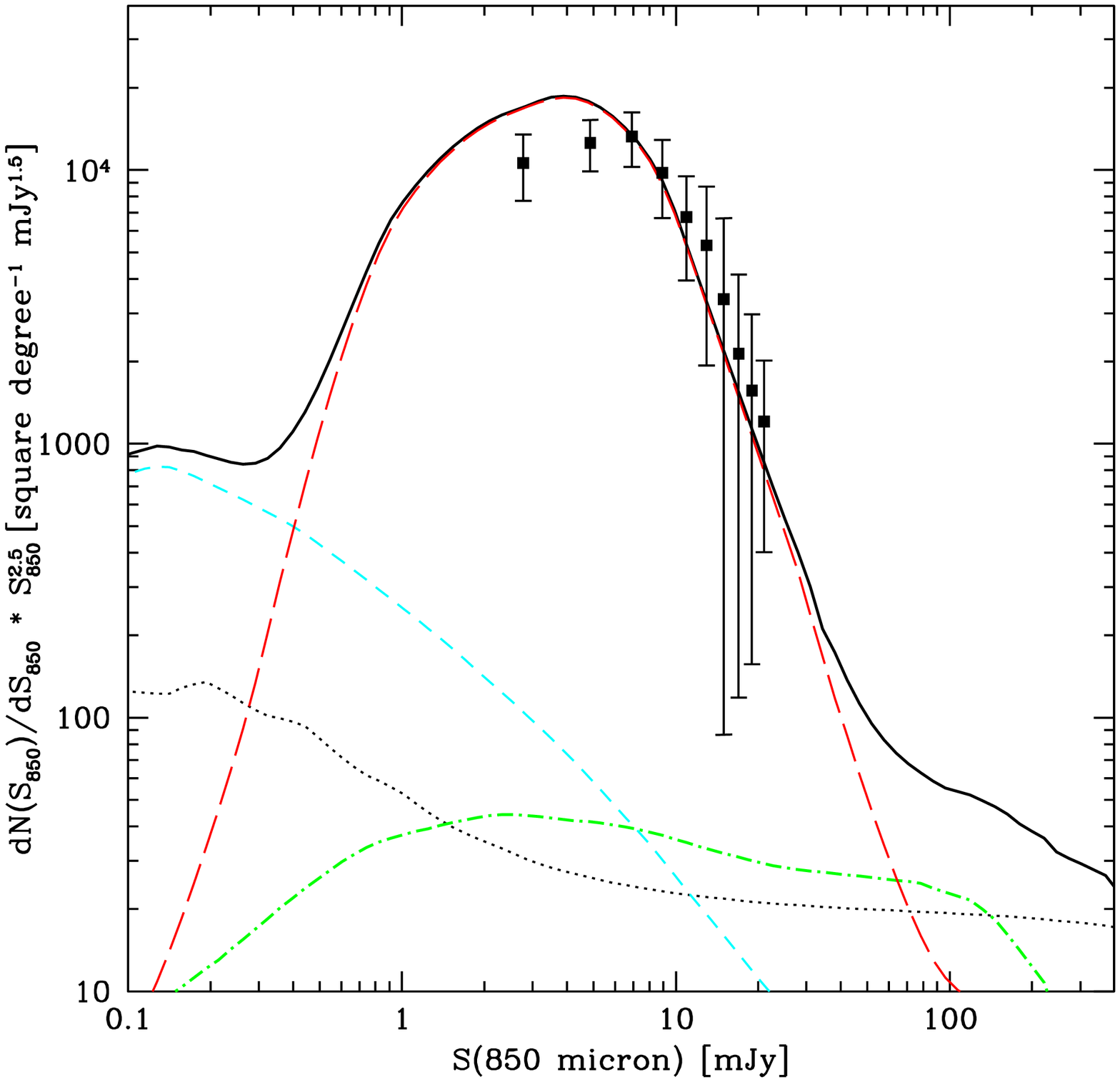}} 
\end{minipage}
\begin{minipage}{0.38\textheight}
\resizebox{8.2cm}{!}{
\includegraphics{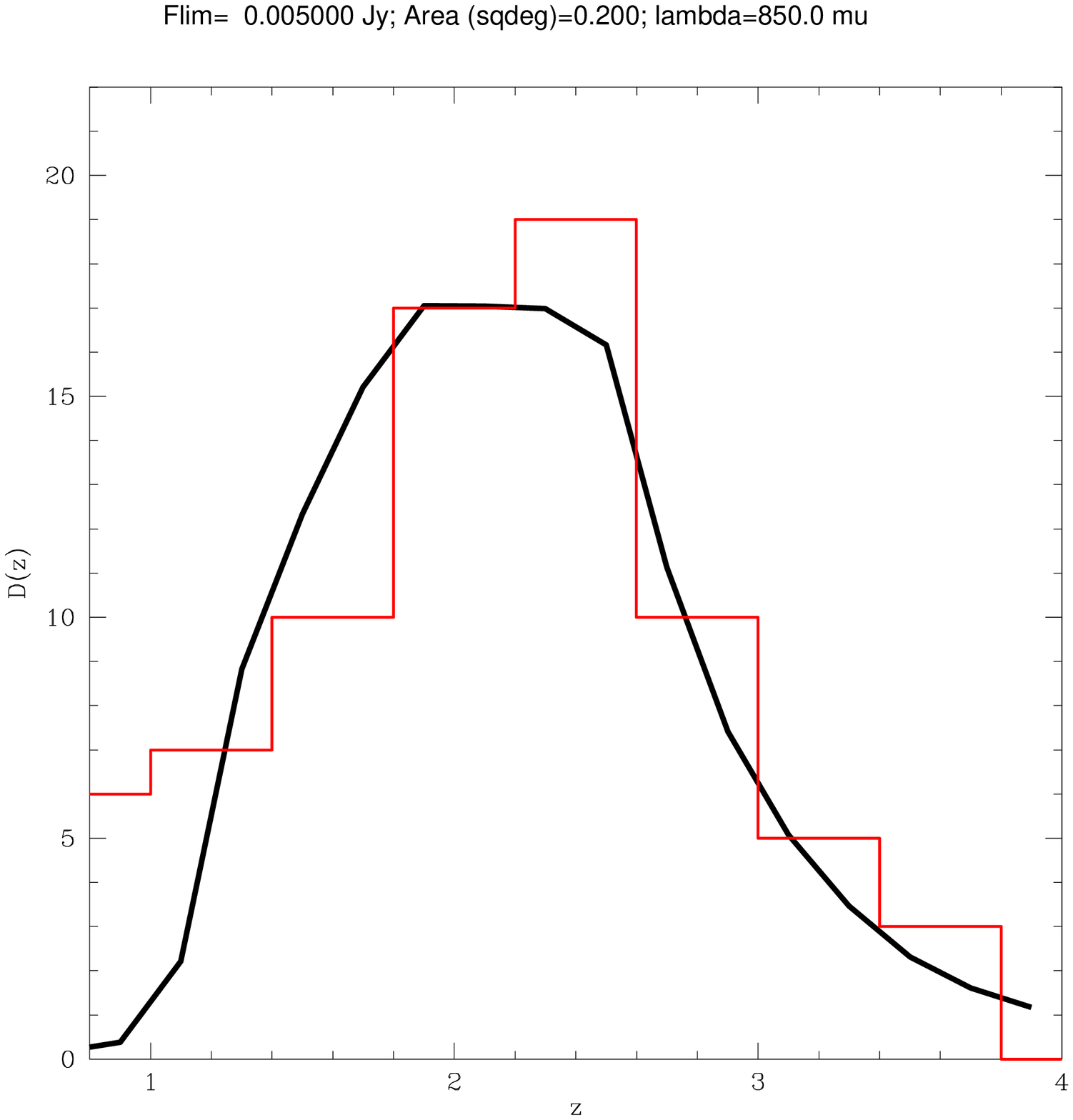}}
\end{minipage}
\caption{Statistics of SCUBA sub-mm sources.
\textit{Left: } The Euclidean-normalized differential counts by Coppin et al. (2006) compared with the model expectation. 
\textit{Right: } The observed redshift distributions of 850 $\mu$m sources reported by Chapman et al. (2005, red histogram), compared with our model prediction for $S_{850}> 5$ mJy (arbitrary normalization).  The datapoint at z=1.6 has been corrected to account for incompleteness in the spectroscopic survey.
 } 
\label{dz850}
\end{figure*}

Vaccari et al. (20010, and in preparation) have considered all three SWIRE MIPS catalogues at 24, 70, and 160 $\mu$m and performed a completeness analysis of the three samples using Monte Carlo simulations.
A likelihood ratio algorithm (Vaccari et al. 2008) was used to perform a cross-correlation of the SWIRE and SDSS samples. We also made use of the IRAC survey in the whole field to eliminate spurious or uncertain identifications. The extension of the area covered by IRAC (9 sq.deg.) set the size of the combined survey. 

Eventually, the compromise MIPS flux limits have been set at 1.5, 20, and 90 mJy at 24/70/160 $\mu$m, respectively, values that ensure a high level of completeness. The redshift boundary for the computation of the local luminosity function was set in all cases to $z_{m}=0.2$, that guarantees good statistics and dynamic range in luminosity and completeness in the redshift information. For $z<z_{m}$ the fraction of spectroscopic redshifts is still large ($>50\%$) compared to the less secure photometric ones. With these boundaries, our final samples included 1058, 664, and 293 sources in the three MIPS channels at 24, 70, and 160 $\mu$m, respectively.

The local luminosity functions were calculated using the 1/V$_{max}$ estimator and Poisson error statistics. The K-corrections have been simply obtained with power-law fits to the average spectra in the narrow frequency interval defined by $z_{m}$, with flux densities assumed to scale as $S_\nu\propto \nu^{-2.5}$ at 24 and 70 $\mu$m and constant with frequency at 160 $\mu$m.

Preliminary results of this analysis are reported in Figs. \ref{llf24}, \ref{llf70}, and \ref{llf160}. Our results in Fig. \ref{llf70}, in particular, are compared with those obtained from the IRAS all-sky survey and later spectroscopic follow-up by Sauders et al. (1990). This comparison shows a fairly good match of the two, except in our highest luminosity bin, where our statistics are poor compared to IRAS.
The same figures also report modellistic fits that will be discussed later.

\begin{figure*}[!ht]
\centering
\includegraphics[angle=90,width=0.95\textwidth,height=0.65\textwidth]{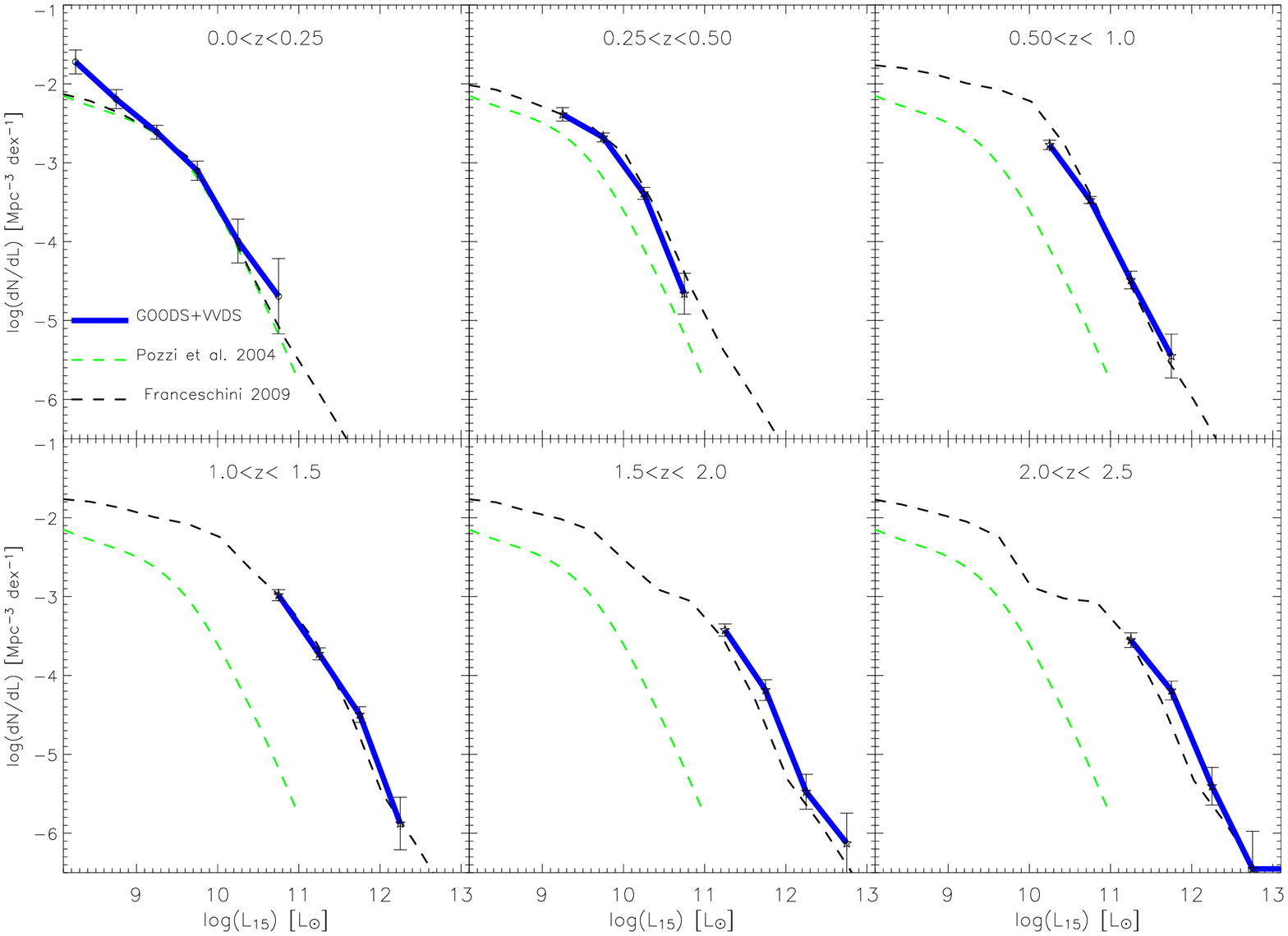}
\caption{Galaxy LF's in the rest-frame 15 $\mu$m in six redshift intervals.
The blue thick lines indicate the luminosity functions obtained by Rodighiero et al. (2010) in their combined GOODS+VVDS-SWIRE sample.  The thick green dashed lines are the local LF by Pozzi et al. (2004) and are reported in all panels for comparison. The upper black dashed lines are our best-fit model prediction. 
}
\label{LF15}
\end{figure*}

\begin{figure*}[!ht]
\centering
\includegraphics[angle=90,width=0.95\textwidth,height=0.65\textwidth]{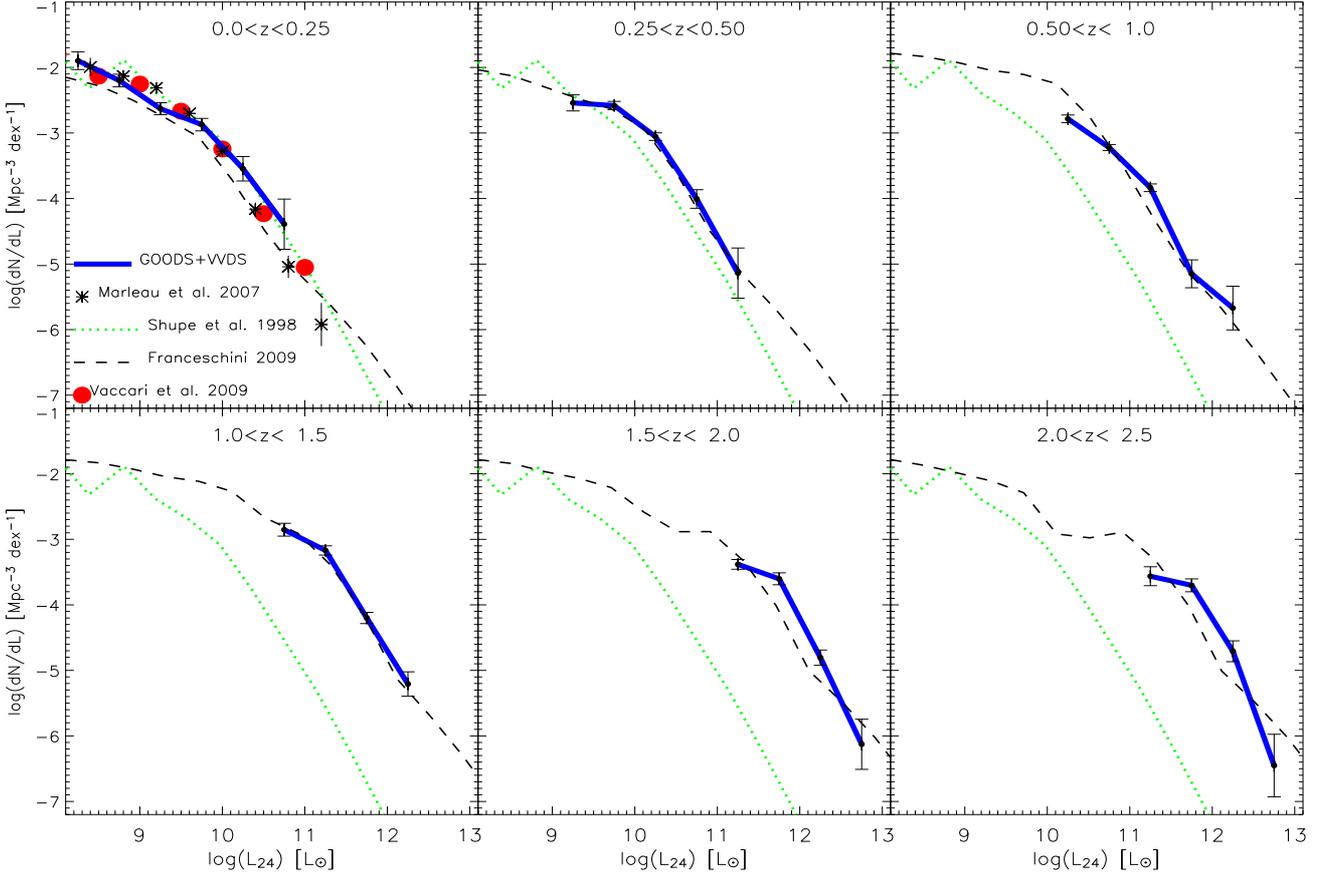}
\caption{Galaxy LF's at 24 $\mu$m in six redshift intervals, (same as Fig. \ref{LF15}, except at 24 $\mu$m). In the low-redshft bin, a comparison is performed with results from Vaccari et al. (2009 in preparation), Marleau et al. (2007) and Shupe et al. (1998).  The latter is used in all panels to show the effect of redshift evolution.
}
\label{LF24}
\end{figure*}

\begin{figure*}[!ht]
\centering
\includegraphics[angle=0,width=0.8\textwidth,height=0.65\textwidth]{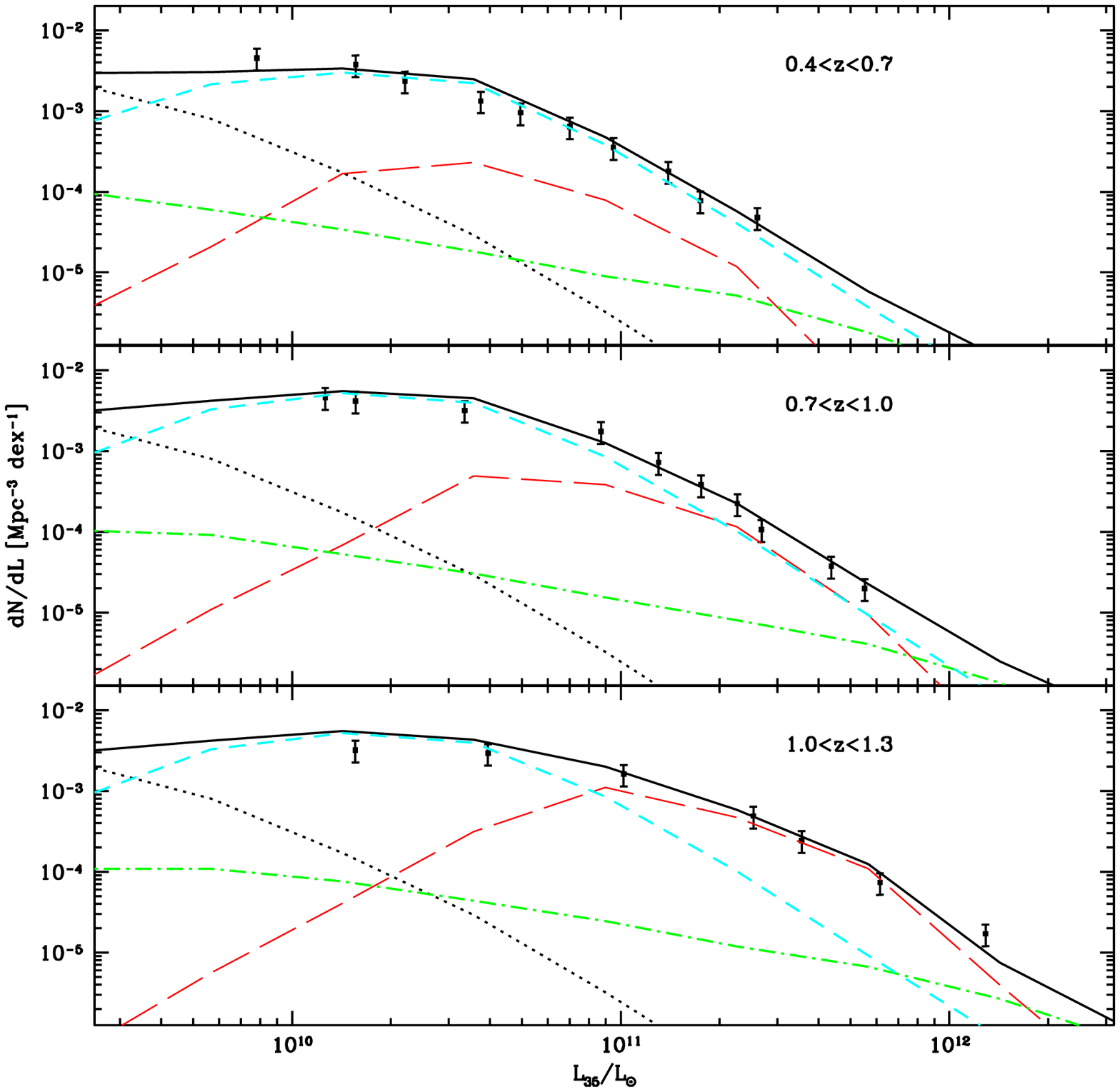}
\caption{Galaxy LF's at 35 $\mu$m in three redshift bins, based on the FIDEL 70 $\mu$m selected sample by Magnelli et al. (2009). 
}
\label{LF35}
\end{figure*}

\subsection{The redshift distributions for faint IR sources}

Thanks to \textit{Spitzer}'s MIPS sensitivity, faint source catalogues from its surveys include substantial numbers of high and very high redshift objects. This is particularly the case for the 24 $\mu$m catalogues from GOODS, SWIRE, FLS, COSMOS, and other surveys, while those at 70 and 160 $\mu$m achieve shallower depths and suffer from the more problematic optical identification. For the first time with \textit{Spitzer}, we can infer relevant constraints on the galaxy and AGN infrared emissivity far above redshift of unity, including the main assembly phases of galaxies and of the cosmological structure.

The simplest and most direct observational constraint on the high-redshift evolution can be inferred from the distribution of the redshifts for complete flux-limited source samples. The procedure to obtain this simple information rests on the optical identification of all sources in the sample and on a measurement of their (either spectroscopic or photometric) redshift. Once this is obtained, a comparison with the modellistic expectation for the redshift distribution is easily done via a simple integral of the redshift-dependent LF over luminosity. 

Rodighiero et al. (2010) have analysed multi-wavelength data, including deep photometric imaging with \textit{Spitzer}, in three areas -- GOODS-South, GOODS-North and SWIRE/VVDS -- and have derived flux-limited samples at 24 $\mu$m. The sky areas are 168, 180, and 2160 square arcmin, and the limiting fluxes 80, 80, and 400 $\mu$Jy, respectively. The corresponding final number of sources in the complete catalogues are 666, 889, and 1495 for GOODS-S, GOODS-N and SWIRE/VVDS, of which 64\%, 33\%, and 9\% have spectroscopic redshifts. The remaining sources had reliable redshift estimates from photometric analyses, as explained in Rodighiero et al. 
These \textit{Spitzer} 24 $\mu$m selected samples serve also as a basic dataset for the luminosity function studies in the next Section.

The optical identifications of these 24 $\mu$m source samples were performed on the basis of a maximum-likelihood technique taking the chance for spurious matches into account in detail. This procedure was particularly successful thanks to the availability of very deep imaging with Spitzer/IRAC in the near-IR in all fields, providing a spectral bridge between the far-IR and the optical catalogues. As a result, only minor fractions of the \textit{Spitzer} sources lacked an optical association. They were 3\% and 1.7\% in GOODS-S and GOODS-N, while essentially all sources in SWIRE/VVDS were reliably identified given the moderate depth of the primary IR selection.
Fig. \ref{dz24} summarizes our results for the redshift distributions in the deepest areas, compared with our model predictions, for two flux limits.   

On the side of sub-mm selected source samples, with their optical identification and spectroscopic characterization made so difficult by the optical faintness and large error boxes, the most systematic attempt to determine their redshift distribution is reported in Chapman et al. (2005) based on a combined SCUBA and very deep radio survey. 
The right-hand panel in Fig. \ref{dz850} shows a comparison of the Chapman et al. redshift distribution with our model's expectation.

\subsection{Direct estimates of the far-IR luminosity functions in redshift intervals}
\label{ir24}

Ideally, quite detailed statistical information on galaxy emissivity and its cosmological evolution can be inferred from redshift surveys of complete flux-limited samples, like those described in the previous section. A popular procedure is to derive estimates of the galaxy luminosity functions in redshift intervals. In practice, this approach suffers various limitations. There is a first obvious need for very large source samples if one wants to achieve an acceptable resolution and to gather enough statistics in redshift space. More seriously, complete samples with (either spectroscopic or photometric) redshift information require not only an enormous effort of spectroscopic follow-up, but also ultra-deep imaging in complementary optical and near-IR wavelengths to obtain reliable estimates of photometric redshifts for the - typically large - fraction of sources without any spectroscopic detection. The implication is that a complete redshift coverage is only achieved over small sky areas and moderate depths, and appropriate sampling of the luminosity function is not possible at low luminosity, except for very low-redshift bins.
Finally, a representative sampling of the whole redshift-luminosity observable plane would require combining ultradeep samples over small sky areas with shallower surveys on much larger fields (the so-called \textit{wedding-cake} cosmological surveying).

Attempts to reconstruct the galaxy and AGN far-IR luminosity functions by various authors (e.g. Babbedge et al. 2007; Caputi et al. 2006, 2007) have exploited the extraordinary deep-mapping capabilities of Spitzer/MIPS at 24 $\mu$m. We made use of the results of a systematic analysis carried out by Rodighiero et al. (2010) on the basis of the optimal combination of depths and areas offered by the three \textit{Spitzer} samples described in the previous section.

While we defer to Rodighiero et al. for any detailed information, we report the results of that work in six redshift bins from z=0 to 2.5 in Figs. \ref{LF15} and \ref{LF24}, based on the combined GOODS-S, GOODS-N and SWIRE/VVDS sample including a grand-total of 3050 galaxies.
Figs. \ref{LF15} and \ref{LF24} illustrate the data-to-model comparisons at 15 and 24 $\mu$m, respectively.  In both figures the effects of the flux limits in the various surveys are apparent, preventing detections of faint sources at progressively higher redshifts. If sources down to low luminosities ($L\simeq 10^8\ L_\odot$) are detectable in the lowest redshift interval, this threshold luminosity increases to $10^{10}\ L_\odot$ by z=1 and $10^{11}\ L_\odot$ at z=2.5. 
We defer to Rodighiero et al. for an extensive comparison with other published results in the literature, showing a good overall agreement.

Overall, there appears to be excellent agreement between the model and data, particularly at $\lambda_{eff}=15\ \mu$m where the K-correction for the data is minimal at the typical source redshift of $z\approx1$. The match is slightly less precise at $\lambda_{eff}=24\ \mu$m, particularly at high redshifts where K-corrections are substantial.

Another aspect to note in the top-left panel of Fig. \ref{LF24}, corresponding to the local LF from the combined sample, is an excellent match with the results of the SWIRE/SDSS analysis in Sect. \ref{llf} and Fig. \ref{llf24}. The two datasets  seem to be in slight excess compared with the model, perhaps due to some large-scale inhomogeneities in the small local volume sampled.

We also compare in Fig. \ref{LF35} our model's results with the luminosity functions in redshift bins based on the FIDEL sample by Magnelli et al. (2009). This sample is based on very deep maps with Spitzer/MIPS in GOODS-N, GOODS-S and ECDS and includes 680 sources selected at 70 $\mu$m with almost complete photometric or spectroscopic redshift information. As apparent in the figure, excellent agreement with the model is found in this case too.

\section{Multi-wavelength modelling of IR galaxy evolution}
\label{model}

Our adopted multi-wavelength modelling of the IR galaxy evolution follows a similar approach to the one pursued by Franceschini et al. (2001, AF2001), which was calibrated on a much shallower database essentially restricted to the ISO survey data. Compared with that scheme, some significant improvements were required, as discussed later.

The present multi-wavelength analysis is not intended to provide us with a detailed and complete justification of the complex physical processes involved in star-formation and galaxy assembly and traced by the IR emission. Instead the model guarantees the best possible adherence to an enormous variety of data on the cosmic source emission from the local to the distant universe and is intended to identify general patterns in galaxy evolution useful for comparison with more physically-based galaxy formation models.

\subsection{The population composition}
\label{population}

Our model population includes contributions by various source categories characterized by different physical and evolutionary properties.

   \begin{enumerate}

\item
The first considered class of sources are non-evolving normal spirals that dominate the multi-wavelength LLF's at low luminosities. Such a population was required in the AF2001 analysis to explain some features in the extragalactic source number counts, particularly those at 15$\mu$m available at that time: an Euclidean slope of the counts down to the milli-Jy level, with a sudden super-Euclidean excess appearing at fainter fluxes. This behaviour is completely confirmed by the new Spitzer/MIPS 24$\mu$m counts shown in Fig. \ref{c24} and is also apparent in the \textit{Spitzer} data at 160 $\mu$m and partly those at 70 $\mu$m in Figs. \ref{c160} and \ref{c70}. 
These data are inconsistent with the whole local galaxy population to evolve back in cosmic time and instead they require the summed contributions of two source categories, a non-evolving and a fast-evolving population. We identify this non-evolving component as a population of spiral galaxies, building up progressively with cosmic time, as indicated by the rather flat age distributions of their stellar populations observed in local objects. We assumed for this class of sources a redshift of formation of $z_{form}=1$ and a local normalization consistent with the observed local luminosity functions discussed in Sect. 2.
Our adopted IR spectral energy distributions (SEDs) for spiral galaxies are discussed in Sect. \ref{spectral}.

\item
A second obvious class of sources to be considered in the model and characterized by high rates of evolution are active galactic nuclei, which are found in substantial numbers in IR selected surveys at all wavelengths (Franceschini et al. 2005; Rowan-Robinson et al. 2005). Among these, type-I AGNs are straightforward to identify based on simple combinations of optical-IR colours (Hatziminaoglou et al. 2005; Siana et al. 2008). For the complementary category of type-II objects instead, the identification would at least require the combination of far-IR/sub-mm data with very deep optical, near- and mid-IR imaging to detect the very hot circumnuclear dust emission characterizing them. For this and other reasons, this class of sources will be treated separately later. 

The local luminosity function for type-I AGNs has been taken from AF2001, which was essentially fitting data by Rush et al. (1995) based on the IRAS 12 $\mu$m survey.
Our assumed LLF, shown as green dot-dashed line in Fig. \ref{llf12} and the following, is entirely consistent with the Rush et al. Seyfert-1 LLF. The LLFs at other IR wavelengths are also reported in Figs. \ref{llf24}, \ref{llf70}, \ref{llf160}.

Type-I AGNs and quasars are modelled based on simple evolutionary recipes inferred from optical and X-ray quasar surveys and with standard UV-optical-IR-mm spectra. As a guideline for assessing their evolution properties in the infrared, we have used the catalogue of 1493 SWIRE sources selected at 24 $\mu$m in an area of 0.85 sq. degree in SWIRE/VVDS to a flux limit of $S_{24}=400\ \mu$Jy as reported by Rodighiero et al. (2010, see Sect. \ref{ir24}). A spectral analysis making use of the optical and IR data available in the field for all objects in the complete sample is reported by Rodighiero et al., among which the type-I AGN (that is quasars and Seyfert-1 galaxies) are easily identified by their flat spectral shapes over the optical through IR wavelength interval. About 8.3\% of the sources at the flux limit are found to be type-I QSO or Seyferts.
We have used the 24 $\mu$m number counts from this catalogue to constrain the evolution properties of Type-I AGNs in our model. In addition, the bright ends (where AGN contribute significantly) of the redshift-dependent luminosity functions at 24 and 35 $\mu$m discussed in Sect. \ref{ir24}, are used to further constrain this evolution.
Based on these data, type-I AGNs are assumed to evolve in luminosity as $L(z)=L(0)\times (1+z)^{3.6}$ up to $z=2$, with $L(z)=L(0)\times 3^{3.6}$ above.
At the same time, these data indicate for the population a moderate comoving density evolution like $dN/dL\propto (1+z)^{0.3}$ up to z=2.

\item
As discussed by AF2001 and Fadda et al. (2002) among others, AGNs of any kinds are far insufficient to explain the faint IR source statistics, because their high IR luminosities contribute to the counts only at rather bright fluxes ($S_{24}\geq 10\ mJy$), while they are too rare to explain the large observed numbers at fainter limits. To try explaining the fainter number counts we need to refer to a much more numerous population of fast evolving objects. 
A third component was then considered, starburst galaxies of moderate luminosities, or LIRGs, which are already present in AF2001, Chary \& Elbaz (2001), Elbaz et al. (2002), Eales et al. (1999), Lagache et al. (2003) and Takeuchi et (2006), among others. 
The increasing relevance of this population with redshift has been one of the main outcomes of the ISO mission and has been confirmed by the Spitzer/MIPS observations (Le Floc'h et al. 2005; Perez-Gonzales et al. 2005). All these analyses found that the MIPS data require a fast increase in the comoving emissivity up to at least $z\simeq 1$. 

The local fraction of the evolving moderately-luminous starburst population is assumed to be $\sim 10$ percent of the total galaxy population (see Fig. \ref{llf12}), roughly consistent with the local observed fraction of interacting galaxies.
The quick observed upturns in the normalized differential number counts in Figs. \ref{c24} to \ref{mm} then require a strong increase in the average emissivity of the evolving population, an increase with redshift that is obtained with both luminosity and number density evolution:
\[   \frac{dN}{dL}(L_{12}[z],z) = \left(\frac{dN}{dL}\right)_0(L_{12})\times (1+z)^{5.2} \ \  \ \ \ z<z_1  \]
\[  \frac{dN}{dL}(L_{12}[z],z) = \left(\frac{dN}{dL}\right)_0(L_{12})\times (1+z_1)^{5.2} \ \  \ \ \ z_1<z<z_{max}  \]
\[ L_{12}(z) = L_{12}\times (1+z)^{2.9}   \ \  \ \ \ z<z_2  \]
 \begin{equation}
 L_{12}(z) = L_{12}\times (1+z_2)^{2.9}   \ \  \ \ \ z_2<z<z_3
\label{solu}
\end{equation}
where $\frac{dN}{dL}(L_{12}[z],z)$ is the redshift-dependent luminosity function at our reference wavelength of 12 $\mu$m, and where $z_1=0.72$, $z_2=0.63$ and $z_3=1.2$. For $z>z_3$ the average population luminosity converges like $L_{12}(z) \propto (1+z)^{-3}$, up to $z_{max}=5$.
This detailed pattern for the evolution of moderate-luminosity starbursts will be discussed later in comparison with that of higher-luminosity sources.

\item
Current data impose an important modification to the AF2001 evolutionary scheme: they require the introduction of a fourth evolutionary population component of very luminous starburst galaxies, or ultra-luminous infrared galaxies (ULIRGs), dominating the cosmic IR emissivity above $z\simeq 1.5$. This latter emerges in particular from the analysis of the galaxy samples selected by the deep Spitzer/MIPS 24 $\mu$m surveys, and their improved capability of detecting cosmic far-IR sources at redshifts $z>1$ and up to $z\simeq 3$ (see Fig. \ref{spectrum} below). As discussed later in Sect. \ref{discussion}, these observations have revealed the existence of a separate population of ULIRGs at high-redshift, essentially absent or very rare locally, hence characterized by an extremely fast evolution in cosmic time. Millimetric observations with SCUBA/JCMT since several years, confirmed by other more recent observations, have also found clear evidence of such very luminous objects at high redshifts (see Figs. \ref{dz24} and \ref{dz850}). 

In our interpretative scheme the distinction between LIRGs and ULIRGs is not to be considered as literally related with their luminosities at any redshifts. Indeed, as shown in Fig. \ref{LFbol} below, at low redshifts their luminosity distributions are similar, but they become quite distinct at $z>1$, where LIRGs have peak luminosity around $L_{bol}=10^{11}\ \L_\odot$, and ULIRGs around $L_{bol}=10^{12}\ \L_\odot$.

Our modelling of this population assumes that it makes a negligible contribution to the local LF's (shown as red long dash lines in Figs. \ref{llf12} to \ref{llf160}), mainly confined to the highl luminosity end of the LLF's, and then evolve in time in a similar fashion as eq. \ref{solu}:
\[   \frac{dN}{dL}(L_{12}[z],z) = \left(\frac{dN}{dL}\right)_0(L_{12})\times (1+z)^{4.1} \ \  \ \ \ z<z_1  \]
\[  \frac{dN}{dL}(L_{12}[z],z) =  \left(\frac{dN}{dL}\right)_0(L_{12})\times (1+z_1)^{4.1} \ \  \ \ \ z_1<z<z_{max}  \]
\[ L_{12}(z) = L_{12}\times (1+z)^{3.9}   \ \  \ \ \ z<z_2  \]
 \begin{equation}
 L_{12}(z) = L_{12}\times (1+z_2)^{3.9}   \ \  \ \ \ z_2<z<z_3
\label{solu2}
\end{equation}
where $\frac{dN}{dL}(L_{12}[z],z)$ is the redshift-dependent ULIRG luminosity function at our reference wavelength of 12 $\mu$m, and where $z_1=1.3$, $z_2=1.3$ and $z_3=1.2$. For $z>z_3$ the average population luminosity converges like $L_{12}(z) \propto (1+z)^{-2.9}$.

   \end{enumerate}

Our assumption here is to include in the two starburst categories also type-II AGNs of moderate or high IR luminosity (type-II Seyfert galaxies and dusty quasars), whose properties of nuclear emission are always associated with various levels of circum-nuclear starburst activity. Disentangling the two physical emissions, particularly in faint distant sources, is a major problem of current-day cosmology (see e.g. Daddi et al. 2007 for \textit{Spitzer} sources and Alexander et al. 2005 for SCUBA objects). In addition to a close relationship of starburst and dust-obscured AGN emission observed in type-II AGNs, there is also evidence of a similar rate of cosmological evolution at least up to redshift z=1 for the two source categories (e.g. Franceschini et al. 1999). Finally, the IR spectra of starbursts and obscured AGNs reveal rather similar shapes (e.g. Berta et al. 2003; Polletta et al. 2007), while differences amount to a slight excess mid-IR emission in dusty AGNs. For all these reasons, and in the spirit of keeping the model complexity and number of free parameters to a minimum, we included type-II AGNs into the starburst categories.

The AGN contribution to the LIRG and ULIRG classes is a major question discussed in Sects. \ref{origin} and \ref{SFRH} below.

\begin{figure*}[!ht]
\centering
\includegraphics[angle=0,width=0.75\textwidth,height=0.55\textwidth]{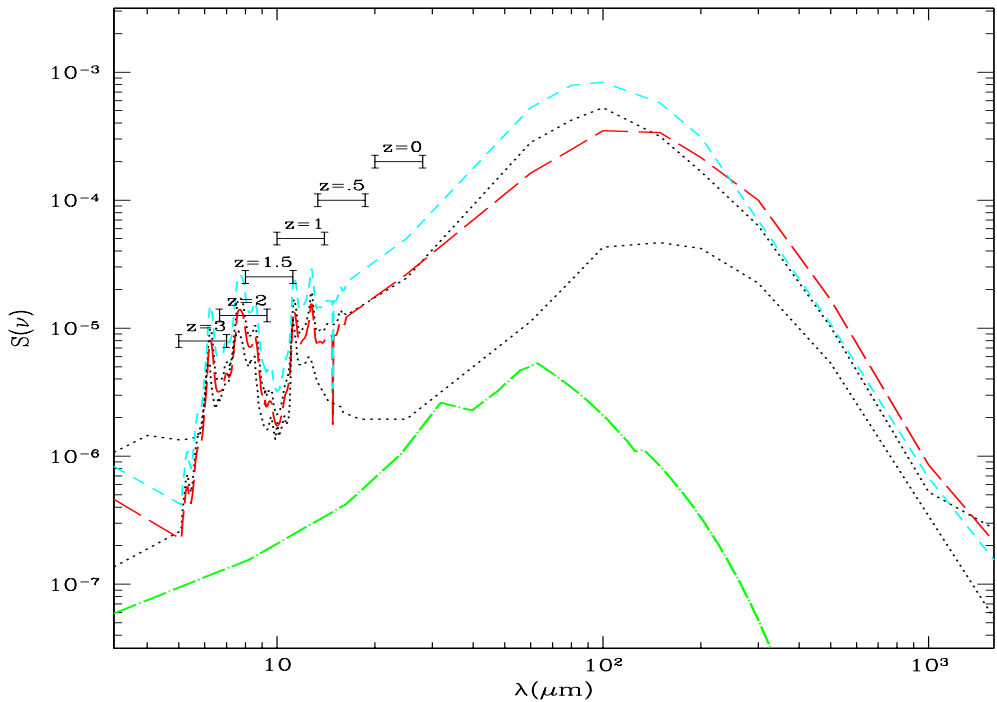}  
\caption{Our adopted IR spectra of various galaxy populations.  The short-dashed cyan line corresponds to our adopted spectrum for the moderate-luminosity LIRG starburst population, while the red long-dashed curve is the spectrum of high-luminosity ULIRG sources. In both cases the spectra are similar to that of the prototype star-forming galaxy M82 (in the range from 5 to 18 $\mu m$ it is precisely the ISOCAM CVF spectrum of M82). 
The lower dotted line corresponds to a low-luminosity inactive spiral ($S_{low}[\nu]$ in eq. \ref{S}), while the upper dotted line is closer to that of ULIRGs ($S_{high}[\nu]$).
The lower dot-dash green line is the average type-I AGN spectrum. The boundaries of the MIPS 24 $\mu$m filter are also shown in the source rest-frames at various source redshifts: due to the prominent PAH features, we expect relative maxima in the redshift distributions at $z\sim 1$ and $\sim 2$, in agreement with the observations (Fig. \ref{dz24}).
}
\label{spectrum}
\end{figure*}

\subsection{The spectral model}
\label{spectral}

As anticipated in Sect. \ref{llf} and following AF2001, our multi-wavelength modelling scheme assumes as a basic reference the local luminosity functions of various populations based on the IRAS 12 $\mu$m all-sky survey. This choice is justified by the fact that the IRAS survey was, and still is, the only all-sky IR survey for which a complete source identification and spectroscopic follow-up has been obtained, allowing us a reliable estimate of the galaxy local IR luminosity functions.
   Fig. \ref{llf12} shows a comparison of our modellistic LLF and the observed datapoints at the closeby effective wavelength of 15 $\mu$m, showing good agreement of the two. From this LLF, luminosity functions at all IR wavelengths are then computed based on a broad-band spectral model.
As for the latter, our choice was again to keep to a minimum the complexity of the model. 

In consideration of the multi-wavelength nature of our used database, the definition of a spectral model is particularly important. Our choice has been to define \textit{a-priory} only the average spectrum of the type-I AGNs, because this has little influence on the model outcomes except for the 24 $\mu$m statistics, as previously discussed in Sect. \ref{population}. For the other population components (spirals, LIRGs and ULIRGs, with the exception of the low-luminosity spirals which have cooler spectra), we have basically referred our model spectra to that of the prototype local starburst galaxy, M82. We have then made adjustments to it so as to achieve a best-fitting to the available data from the mid-IR throughout the millimetre. We anticipate that an interesting outcome of our analysis was that small modifications to the M82 broad-band spectrum are sufficient for obtaining this best-fit.

\subsubsection{AGN}

For the type-I AGN population we assume an SED corresponding to an emission model by a face-on dusty torus and is reported as a green dot-dashed line in Fig. \ref{spectrum}. This average spectrum is taken from the radiative transfer model by Fritz et al. (2006), and was optimized to fit photometric data for AGNs in the SWIRE surveys (see Franceschini et al. 2005; Hatziminaoglou et al. 2008). 

Our adopted type-I AGN spectrum is close to and consistent with the average Spitzer/IRS spectral shapes of quasars published by Netzer et al. (2007). A discussion is reported there about the exact origin of the far-IR flux of quasars and an indication is provided that this long-wavelength part, as opposed to the mid-IR spectrum which is certainly due to AGN circum-nuclear dust, might instead be due to stellar-illuminated dust. For our present purposes it makes no much difference if part of the AGN population radiation is due to starburst rather than gravitational processes.
Note finally that our detailed spectral modelling for AGNs is not critical, because type-I AGNs contribute significantly only to the 24 $\mu$m, and to a lower extent to the 70 $\mu$m, statistics at the bright fluxes.

\subsubsection{Spirals}

Also in the case of the normal spiral population a precise definition of the spectral model is not critical for our attempt to interprete the high-redshift source evolutionary properties. Although spirals  dominate the low-luminosity end of the LLFs, their lack of evolution implies that their number density becomes negligible compared to other evolving populations at cosmological redshifts. 
As discussed in AF2001, a comparison of LLF's in the mid-IR, far-IR and sub-mm shows that the 60$\mu m$ LLF has a flatter power-law shape at high-luminosities compared with the other LLF's, an effect of the spectra for more luminous galaxies showing excess 60 $\mu m$ emission compared with inactive galaxies (see also the luminosity-dependence of the IRAS far-IR colours of galaxies; Soifer et al. 1989; Chary \& Elbaz 2001). We have then modelled the redshift-dependent multi-wavelength LLF's of spiral galaxies by assuming SEDs dependent on luminosity, with spectra ranging from those typical of low-luminosity inactive objects with peak emission at 200-300 $\mu$m, to those peaking at 60-80 $\mu$m characteristic of luminous luminous IR galaxies. 

We took as reference for our multi-wavelength LF the one by IRAS at 12 $\mu m$ ($\frac{dN}{dL}(L_{12})$) discussed in Sect. 5.1. 
This is transformed to longer wavelengths
according to SEDs that vary according to the value of $L_{12}$.
The assumption was that for $L_{12}<2\ 10^9\ L_\odot$ the spectrum is that of an inactive spiral  ($S_{low}[\nu]$, the lower black dotted line in Fig. \ref{spectrum}), while for $L_{12}>10^{12}\ L_\odot$ it is a typical ULIRG spectrum ($S_{high}[\nu]$, upper dotted line in Fig. \ref{spectrum}).
For intermediate luminosity objects, the assumed SED, $S(\nu)$, is a linear interpolations between the two as a function of the $L_{12}$ luminosity:
\begin{equation}
S(\nu) |_{L_{12}}={S_{low}(\nu) (\log L_{12}- 9.3) + 
S_{high}(\nu) (\log L_{12}-12) \over 2.7} .
\label{linear}
\end{equation}
The multi-wavelength LLFs at any wavelengths $\lambda$ are then easily computed from the relation
\begin{equation}
\frac{dN}{dL}(L_{\lambda}|L_{12},z) = \frac{dN}{dL}(L_{12},z) \left( d \log L_{\lambda} \over 
d \log L_{12}\right)^{-1}.
\end{equation}
The $S(\nu)|_{L_{12}}$ spectrum from eq. \ref{linear} is also needed to compute the K-correction. 
%

\subsubsection{LIRG and ULIRG galaxies}

Of course, more critical are the assumptions about the spectral fitting of the two luminous galaxy populations considered in the model, the moderate and the high-luminosity objects (i.e. the LIRGs and ULIRGs). To simplify our treatment, we adopted for each one of the two classes a single spectral form, independent of luminosity. For the LIRG and ULIRG objects our best-fit spectra are shown in Fig. \ref{spectrum}. It should be noticed how similar the two spectra are, both of them turning out to be close to the spectrum of the prototype local starburst galaxy M82 (e.g. Polletta et al. 2007). 

We then support earlier findings by AF2001 that an M82-like IR SED is needed to reconcile the various statistics of faint starbursting sources, including those currently detected in large numbers at $z>1$ not sampled by previous ISO observations. In particular, the adoption of the IR SED of the ULIRG galaxy Arp 220, an highly absorbed ultra-luminous source, would imply the far-IR counts and LF's having far too high normalizations compared with the mid-IR ones (at 12, 15, and 24 $\mu$m), and a complete mismatch of the observational data.
The average spectral shapes in Fig. \ref{spectrum} are required to reconcile a vast amount of multi-wavelength data and should then be considered as representative of the bulk of the population.

We stress that, in spite of the simplifying assumption about the average spectra of IR-selected starbursts to be independent of redshift and luminosity,  our model is still able to provide good fits to the various statistical observables. Obviously this is not by itself proof that the IR spectra are completely time- and luminosity-invariant. 


\subsection{Model to data confrontation}

With our spectral model for the different source populations, we computed LFs at all IR wavelengths by convolving the spectra with the detailed filter response functions of the various photometric channels.
Once the redshift-dependent multi-wavelength luminosity functions for our source populations are set, the procedures for comparing them to the variety of data discussed in Sect. 2 are also straightforward. We defer the reader to a detailed account in AF2001 about the computation of  number counts, redshift distributions, etc. Here again our spectral model is needed in order to calculate not only the luminosity functions at various wavelengths, but also the K-corrections in the luminosity-distance relation. All this requires that the average source spectra are convolved with the detailed spectral response functions of the various observational filters, as discussed in AF2001.

As illustrated in Figs. \ref{c24} to \ref{LF35}, our \textit{best-fit} model appears to reproduce a wide variety of multi-wavelength IR data, from redshift-dependent luminosity functions to faint number counts, and providing marginally acceptable fits of the background spectral intensity and cell-to-cell fluctuations. 
A detailed comparison of our model for IR galaxy evolution with the luminosity function data in redshift bins at 15, 24, and 35 $\mu$m (Sect. \ref{ir24}) shows excellent agreement up to the maximum observed redshift of $z=2.5$, with overall acceptable values of the total reduced $\chi^2$ ($\chi_\nu ^2\simeq 1.1$). There is only a slight tendency of the model fit to remain low in the lowest bin at 24 $\mu$m and $0<z<0.25$, probably due to the unaccounted effects of the local large-scale structure in the luminosity function data. In any case, the low-z behaviour of the model was calibrated on a variety of number count data at bright fluxes (including the IRAS all-sky far-IR counts and those from the \textit{Spitzer} SWIRE project), which guarantees excellent control of the local IR universal emissivity.

Compared with published analyses of IR galaxy evolution based on direct estimates of the galaxy luminosity functions (e.g. Perez-Gonzales et al. 2005; Caputi et al. 2007; Rodighiero et al. 2010), our current approach of combining these  with integrated information coming from faint number count and background data allows us to better constrain the luminosity functions over wider redshift and luminosity intervals, including some extrapolations to the low luminosities where they are not directly measurable (see Fig. \ref{Ldens} below for more details about uncertainties in this analysis).

\subsection{Model predictions}

Appendix B is dedicated to discussing predictions for number counts by forthcoming observations, particularly with \textit{Herschel}. 
We also report there some expectations for source confusion limitations with future instrumentation. 

\begin{figure}[!ht]
\centering
\includegraphics[angle=0,width=0.5\textwidth,height=0.45\textwidth]{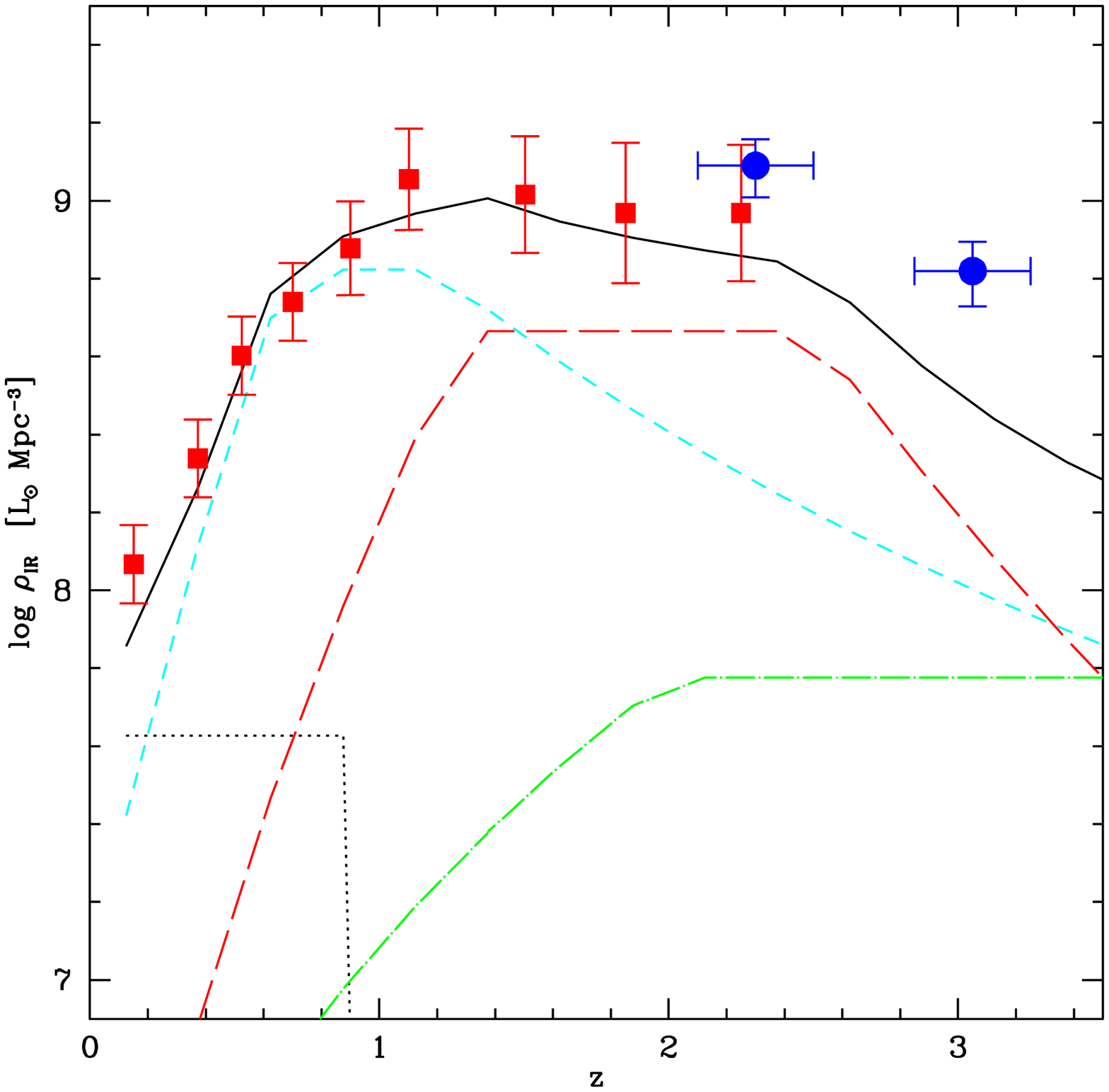}  
\caption{Evolution of the comoving bolometric luminosity density from 6 to 1000 $\mu$m for the IR-selected galaxy population, based on the model discussed in Sect. \ref{model}. The luminosity density is expressed here in solar luminosities per cubic Mpc.
   Green dot-dashed line: type-I AGNs.   Cyan short-dashed line: evolving moderate-luminosity starbursts (LIRG). Red long-dashed line: high-luminosity starbursts (ULIRG). Lower dotted black line: quiescent spiral population (note that the spirals are assumed not to evolve in comoving luminosity by our model and then to disappear at z=1).  The upper continuous line is the total predicted emissivity. 
Red filled square datapoints are from the analysis of the 24 $\mu$m luminosity functions by Rodighiero et al. (2010). Blue datapoints are from a large spectroscopic survey at $1.9<z<3.4$ by Reddy et al. (2008).
}
\label{rho}
\end{figure}

Fits to the observational data and individual population compositions are reported in Figs. \ref{c24} through \ref{rho} with reference to our \textit{best-fit} model.

%

\begin{figure*}[!ht]
\centering
\includegraphics[angle=0,width=0.95\textwidth,height=0.65\textwidth]{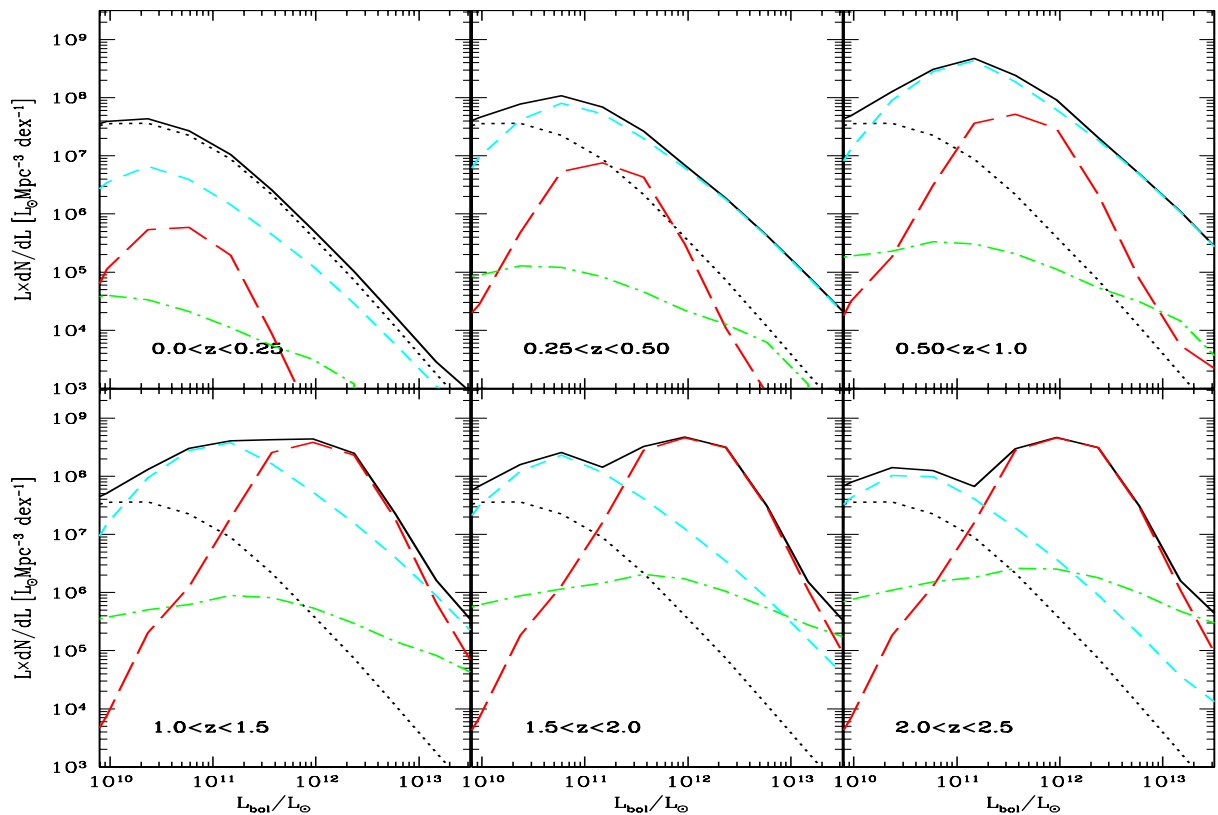}  
\caption{The comoving bolometric emissivity from 6 to 1000 $\mu$m as a function of luminosity for the IR-selected galaxy population, calculated at various cosmic epochs from z=0 to 2.5. The IR emissivity is expressed here in solar luminosities per cubic Mpc per unit logarithmic interval of $L$.
Line types as in Fig. \ref{rho}.
}
\label{LFbol}
\end{figure*}

\section{Discussion}
\label{discussion}

\subsection{General evolutionary patterns for IR sources:  evidence of a strongly luminosity-dependent evolution}
\label{EVOL}

The redshift evolution of the comoving bolometric IR emissivities from 6 to 1000 $\mu$m for our model is reported in Figure \ref{rho}. Individual contributions of our four galaxy classes, the normal spirals, type-I AGNs, moderate-luminosity, and high-luminosity galaxies are also shown. 
Our model predictions are compared with a recent estimate of the IR emission of the global population based on 24 $\mu$m data and on fitting the LF data in redshift bins with two-power-law functions (Rodighiero et al. 2010). Both our model and the data of Rodighiero et al. show a fast increase in the comoving IR emissivity to $z\sim 1$, the total increase for the best-fitting model amounting to a factor $\sim 30$. A flattening of the comoving emissivity is then observed thereafter, while some convergence would be expected by the model at $z>2.5$, as discussed below.

Figure \ref{rho} also reveals some remarkable differences in the evolutionary properties of the two energetically-dominant populations of moderate- (LIRG) and high-luminosity (ULIRG) active galaxies: the epoch of peak activity of the former appears to be shifted to lower redshifts compared to the more luminous ones dominating at the highest redshifts.  Similar findings based on IR source populations are also reported by Le Floch et al. (2005) and Perez-Gonzales et al. (2005).

This general evolutionary pattern is confirmed by the  more detailed analysis reported in Fig. \ref{LFbol}, showing the evolution of the comoving bolometric emissivities from 6 to 1000 $\mu$m as a function of luminosity for the various IR-selected galaxy populations, calculated in six redshift bins from z=0 to 2.5. The IR emissivities on the y-axis are expressed here in solar luminosities per cubic Mpc and per unit logarithmic interval of $L_{bol}$, and are simply given by the product of the bolometric luminosity function times $L_{bol}$ (hence proportional to the contribution of the population to the total emissivity per unit decade of L).

\begin{figure*}[!ht]
\centering
\includegraphics[angle=0,width=0.75\textwidth,height=0.65\textwidth]{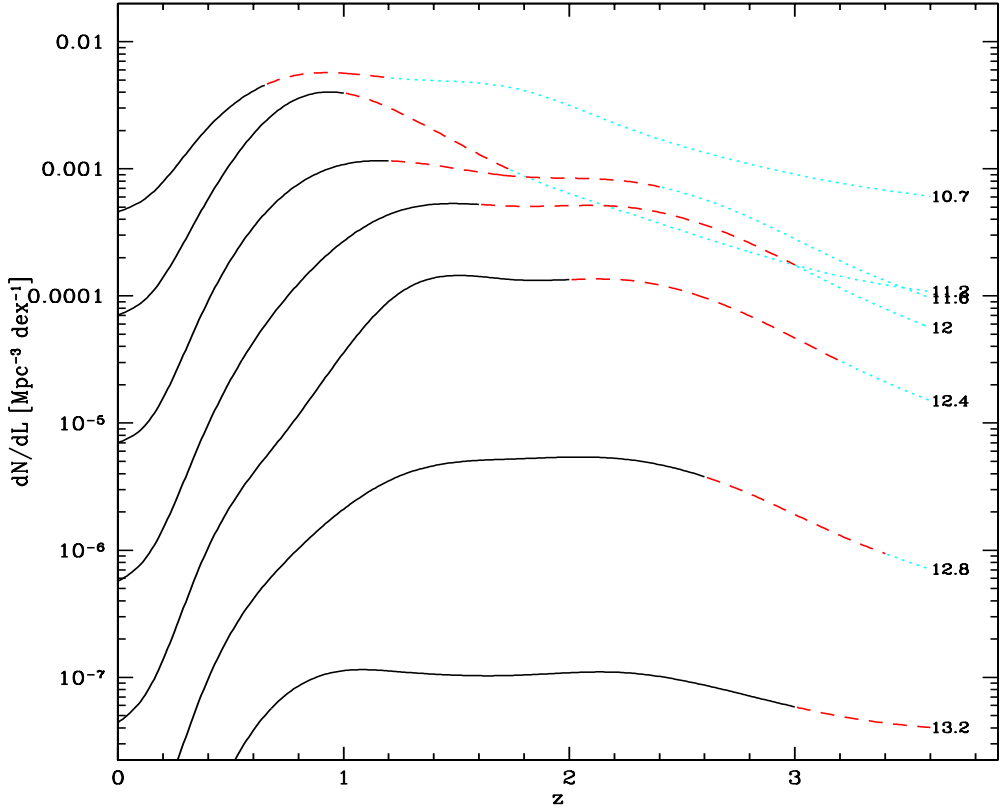}  
\caption{The best-fit model prediction about the comoving number density versus redshift of cosmic IR sources as a function of source luminosity.    The latter, indicated at the right hand side of each curves, is given by $\nu L(\nu)$ at 100 $\mu$m in solar luminosities. 
Line types correspond to redshift intervals with the following coding. Continuous black lines are regimes where the luminosity functions are constrained by current far-IR number counts, particularly the Spitzer 70 $\mu$m counts at the current sensitivity limit, in addition to all other available statistics; the red dashed lines are constrained by the Spitzer 24 $\mu$m counts at the current sensitivity limit (30 $\mu$Jy), 24 $\mu$m luminosity functions and z-distributions; cyan dotted lines are regimes where the IR luminosity functions are constrained essentially by the CIRB spectral intensity. 
As the available constraints decrease going from low- to high-z, the uncertainties in the curves for the various luminosity intervals also increase with redshift.
}\label{Ldens}
\end{figure*}

In the lowest-z bin, the bolometric emission is dominated by inactive spirals, type-I AGNs contribute at the highest-L, while the ultra-luminous starburst ULIRG population provides a negligible contribution at any $L_{bol}$. This situation rapidly changes already by $z\simeq 0.5$, when active galaxies of both LIRG and ULIRG classes start to dominate, except at the lowest values of L.
The moderately-luminous objects (LIRGs) then appear to stop evolving by $z\sim 1$, whereas the ULIRGs continue to rise in both comoving number and bolometric luminosity to higher $z$.

Remarkably enough, the highest redshift bins show a prominent dominance of the ULIRGs, with rather steady comoving number densities as a function of redshift. The LIRG population keeps a bumpy luminosity function at all redshifts, and already appears to decrease in number density and luminosity at $z>1$.

A different view emphasizing the redshift-dependence of the evolution is reported in Fig. \ref{Ldens}, showing the comoving number density of galaxies versus redshift for various luminosity classes, indicated on the right hand side of the figure (these are the average $L_{bol}$ values in the bin).
The figure also includes a detailed assessment of how the main datasets constrain our solution in various luminosity and redshift intervals, thereby providing information about the reliability of our solutions. 
Continuous lines are regimes where the luminosity functions are constrained by current far-IR number counts, particularly the Spitzer 70 $\mu$m counts at the current sensitivity limit, in addition to all other available statistics; the dashed lines are constrained by the Spitzer 24 $\mu$m counts at the current sensitivity limit (30 $\mu$Jy), 24 $\mu$m luminosity functions and z-distributions. Inside these redshift intervals spectral extrapolation to the far-IR is required to estimate the bolometric luminosity, but this is also well constrained by BLAST and preliminary Herschel observations (Sect. \ref{H}). 
Finally, dotted lines are regimes where the IR luminosity functions are constrained essentially by the CIRB spectral intensity and background fluctuations. 
As the available constraints decrease going from low- to high-z, the uncertainties in the curves for the various luminosity intervals also increase with redshift, mostly because of an increase in the degeneracy of the model solutions.

Again interesting trends are shown in Fig. \ref{Ldens}. The rate of increase from $z=0$ to 1 of the comoving density is a strong and monotonic function of luminosity, the rise factor being more than 3 orders of magnitude for the most luminous galaxies, while it is only a factor $\sim 3$ for the faintest ones. (The previously mentioned average factor for the total population is $\sim 30$).
The other main observed feature is a global and systematic shift of the peak emissivity as a function of redshift, the most luminous objects having their maximal activity at $z\geq 2$, while the least luminous ones do it at redshift close to 1.
This details more precisely in terms of the bolometric luminosity class a trend already anticipated in Fig. \ref{rho}.

The lowest luminosity class in Fig. \ref{Ldens} (contributed by spirals and LIRGs) reveals a moderate evolution over the whole redshift range analysed here. It should be noted, however, that the high-redshift behaviour of sources with this luminosity is not reliably sampled by current observations.
For all other luminosity bins with $L_{bol}> 10^{10}\ L_\odot$, instead, the combined dataset of luminosity functions, deep number counts, and background data considered by our analysis provides a relevant set of constraints on the evolution of emissivity of IR galaxies that are not avaliable from galaxy surveys at any other wavelengths.

Finally, the convergence of the comoving emissivity at $z>2.5$ in Fig. \ref{rho} was imposed in particular to avoid exceeding the SCUBA number counts, particularly the two faintest datapoints at $S_{850}$ between 2 and 6 mJy in Fig. \ref{dz850} left, and to a less extent the CIRB COBE-observed intensity at $\lambda>500\ \mu$m. The function $\rho_{IR}$ is dominated here by the ULIRG population. 

The detailed behaviour of the high-z convergence of the ULIRG's comoving volume emissivity (Fig. \ref{rho}) is not yet well constrained by the scanty information on the z-distribution of SCUBA sources (Fig. \ref{dz850}; Chapman et al. 2005).
Equally, the deep Spitzer/MIPS 24 $\mu$m surveys suffer a strong K-correction penalty at $z>3$ (Fig. \ref{dz24}), such that a few Spitzer sources are detected out there. 
 Confirmation, particularly by SCUBA-2, of the source number counts at the faintest levels, and the characterization and redshift distributions of the forthcoming large samples of \textit{Herschel} sources in the sub-mm, will be essential to establish the detailed cosmic evolution at $z>2.5$.
Also later high spatial-resolution studies with ALMA will offer us richer datasets to constrain this early history of galaxy IR emissivity.
We will assume henceforth that $\rho_{IR}$ quickly converges there, as in Fig. \ref{rho}, but will later discuss the uncertainties related with this.

Even when taking this uncertainty into account, there is a clear tendency in our redshift-dependent bolometric comoving energy density shown in Fig. \ref{rho} to stay lower than various published results, particularly at higher redshifts than $z\sim 1.5$.
These include Hopkins (2004), Hopkins and Beacom (2006), and the overview reported in Perez-Gonzalez et al. (2008), and are typically based on either large extinction corrections (for the UV-optical, or H$\alpha$ observations) or large spectral extrapolations (like those making use of the MIPS/24$\mu$m, the SCUBA, or deep radio surveys).
This difference will have important implications for our later investigation.

In conclusion, the evolution of IR emissivity for our best-solution model emphasizes the activity of very luminous objects at the higher redshifts ($z\sim 2.5$), while lower luminosity galaxies appear to dominate at lower redshifts. This kind of evolution is very reminiscent of the \textit{downsizing} scenario that has emerged from studies of the redshift-dependent galaxy stellar mass functions (Bundy et al. 2006; Franceschini et al. 2006; Fontana et al. 2006; Perez-Gonzalez et al. 2009; among others), in which the most massive galaxies and their stellar content are already largely in place at $z>1$, while less massive systems continue to form stars until more recent epochs. 
We seem to have identified a corresponding evolutionary feature in terms of the e.m. (bolometric) emission produced by forming stars (and AGNs).
This also confirms and quantifies evolutionary trends anticipated by Cowie et al. (1996) for galaxies and Franceschini et al. (1999) for both galaxies and AGN (see also Perez-Gonzalez et al. 2008; Mobasher et al. 2009).

However, in terms of bolometric, comoving energy density, this activity appears to decrease at $z>1.5$ and to remain slightly lower than sometimes reported earlier, based on large extrapolations.

\subsection{Origin of the IR emission in the sample sources}
\label{origin}

The big advantage of far-IR and sub-mm observations of faint high-redshift objects is that, particularly for the more luminous source populations, the LIRGs and ULIRGs, these data offer an excellent \textit{proxy} for the bolometric emission (e.g. Soifer et al. 1987; Sanders and Mirabel 1996; Elbaz et al. 2002; Rieke et al. 2009). 
If we know whether the emission is primarily to ascribe to either star-formation or AGN activity, we can then directly infer the evolutionary history of star formation or nuclear gravitational accretion from the IR emissivities.

The problem here is that various different processes tend to mix together in producing IR photons, so that the interpretation may not be trivial. Fortunately, good insight into the physical nature of the primary energy sources in both local and high-redshift IR objects have been obtained with the very sensitive IRS infrared spectrograph on \textit{Spitzer}. Yan et al. (2007), Sajina et al. (2007, 2008), and Lacy et al. (2007), among others, have reported low-resolution spectroscopy in the critical region between 5 and 40 $\mu$m, where the stellar and AGN components show maximal differences.   The IRS spectroscopy of high-redshift IR sources has been possible for the highest luminosity objects, which tend to include a higher incidence of AGNs (Vieilleux et al. 1997; Kim et al. 1998). 

As a general trend, these studies indicate that high-z sources selected in the sub-mm (the SCUBA galaxies) are dominated by starburst emission, while those selected by Spitzer/MIPS in the mid-IR tend to include a higher incidence of AGNs. The relatively higher contribution to the mid-IR than to the sub-mm by very hot circum nuclear dust heated by a nuclear AGN is well established by observational evidence and theoretical expectation. Although including AGN contributions in a large majority of cases, the extensively investigated SCUBA galaxies appear to have their bolometric flux dominated by star formation (Alexander et al. 2005).   Daddi et al. (2007) from the analysis of luminous 24 $\mu$m selected galaxies at $z\sim 2$ find that the majority of the sources more massive than $10^{11}\ M_\odot$ show excess hard X-ray emission in stacked \textit{Chandra} images, revealing trace emission by a Compton-thick obscured AGN.  In general, however, all X-ray studies of the AGN incidence in obscured high-redshift objects have to be taken with caution, due to the fundamental limitation that, above some value of the column density ($N_H \simeq 10^{25}\ cm^{-2}$), the whole X-ray flux is essentially reprocessed to longer wavelengths.
In this regard, X-ray studies provide lower limits to the AGN content.

An interesting analysis of the AGN incidence in IR- and sub-mm-selected sources has recently been reported by Nardini et al. (2008) for local ULIRGs and Watabe et al. (2009) for high-z objects. Their conclusions, based on stacked Spitzer/IRS spectra of samples of sources selected in redshift and luminosity bins, may be summarized as follows. The majority of both sub-mm and 24 $\mu$m sources show evidence of some AGN contribution in the IRS 6 to 8 $\mu$m rest-frame spectra, this fraction becoming dominant for the 24 $\mu$m selected objects. In terms of the bolometric IR-mm luminosity, the AGN contribution appears to mix with what is coming from a starburst. In SCUBA galaxies the overall AGN contribution is estimated at about 7\%, while for the ultra-luminous 24 $\mu$m sources it settles around one third of the bolometric source emission, fairly independently of redshift.  It should be kept in mind that all these results are to be taken as very preliminary: not only are they based on a handful of objects, but the spectral analysis was also performed on the stacked IRS spectra.

Our evolutionary scheme includes a limited ability to disentangle the main different primary power sources, young stars in and around molecular clouds, and AGN emission from gravitational accretion. The type-I broad-lined AGNs are in general easily identied from their flat power-law, mid-IR SEDs and well-characterized optical spectra. Indeed, one of the four source categories in our model are type-I quasars and AGNs. Their local multi-wavelength LF's and evolution rates are well known, and the model predictions have been easily tested on existing data (Sect. \ref{population}).

In contrast, the more numerous type-II AGNs, typically including a mixture of dusty AGN and obscured starburst emissions, are much more confused among what we called the LIRG and ULIRG populations. Because there is no consensus yet on the type-II AGN statistical properties, no attempts have been made by us to differentiate them from the LIRGs and ULIRGs.
The broad-band SEDs of our various classes of sources in our analysis, reported in Fig. \ref{spectrum}, reveal spectral shapes broadly consistent with those of star-forming galaxies, with only type-I AGNs showing the flat and featureless IR spectra of AGNs. 

From all the above, it is expected that a significant fraction of all the IR sources analysed in our paper include some type-II AGN contribution. This should be more frequent among our ULIRG class and more frequent in mid-IR than far-IR and sub-mm samples. For example, Franceschini et al. (2002) suggest that assuming approximately 15\% of the 24 $\mu$m sources to be type-II AGN with bright hard X-ray emission would reconcile statistical properties of IR and X-ray surveys and the CIRB and X-ray cosmological backgrounds.  In fact, it is more appropriate to think of it in terms of a larger majority of all mid-IR selected objects to include AGN components with various degrees of contribution.

We can summarize our current understanding of the origin of the IR emission in cosmic sources with the statement that starburst and obscured AGN activities are found in many cases to occurr simultaneously, with an increased tendency for an AGN to appear at increasing luminosity/mass.  However, the detailed contribution of one or the other energy-production mechanism still remains largely undetermined.  Our next section will provide us with more insight into this problem.


It should be finally noticed that our previously discussed evolution pattern of an increased redshift for the peak activity at increasing luminosity shown in Fig. \ref{Ldens} cannot in any case be biased  much by the AGN component, because type-I X-ray AGNs have revealed very similar tendencies of their comoving number densities vs. redshift as shown here for the whole IR source population (Hasinger, Miyaji and Schmidt 2005; Bongiorno et al. 2008).

%
%
\begin{table*}
\caption{The stellar mass fraction $f_{\ast}(t)$ after gas recycling, as a function of population age for different IMFs of power-law form ($dn/d \log M \propto M^{-m}$), and the $L_{IR}$ to $SFR$ calibration}   
\label{table0}      
\centering                          
\begin{tabular}{ c c c c c c c}        
\hline                 
\hline 
age      & Scalo & Kroupa   & Salpeter  & Top-heavy 1 & Top-heavy 2 & Top-heavy 3 \\
 Gyrs    & (1986)& (1993)   & $m=1.35$  & $m=1.15$  & $m=1.0$   & $m=0.85$   \\    
\hline                                                
  0.01   &   0.98     &   0.97   &   0.90    & 0.87 &   0.83  	&   0.60                \\
  0.05   &   0.94     &   0.92   &   0.85    & 0.74 &   0.62  	&   0.53                \\
  0.1    &   0.92     &   0.90   &   0.83    & 0.70 &   0.58  	&   0.44                \\ 
  1.0    &   0.86     &   0.81   &   0.75    & 0.61 &   0.48  	&   0.35                \\
  3.0    &   0.80     &   0.75   &   0.72    & 0.57 &   0.44  	&   0.32                \\ 
  7.0    &   0.75     &   0.72   &   0.70    & 0.56 &   0.42  	&   0.31                \\
  10.    &   0.73     &   0.70   &   0.65    & 0.55 &   0.41  	&   0.30                \\
  15.    &   0.71     &   0.69   &   0.65    & 0.54 &   0.40  	&   0.30                \\
\hline 
$\frac{SFR\ [M_\odot/yr]}{L_{IR}/L_\odot}$ & $3.3\ 10^{-10}$ & $2.2\ 10^{-10}$ & $1.63\ 10^{-10}$ & $1.01\ 10^{-10}$ & $0.63\ 10^{-10}$ & $0.46\ 10^{-10}$ \\
\hline 
\hline 
\end{tabular}
\end{table*}

\subsection{Matching the average comoving SFR and the integrated stellar mass densities}
\label{match}

Some relevant facts have already emerged from our IR evolutionary model.  
The first is that the very fast increase in the source emissivity from $z=0$ to $\sim 1$ is happening in a population of moderate luminosity galaxies (Fig. \ref{LFbol}, see also Le Floc'h et al. 2005), the LIRGs, with typical luminosities of $L_{bol}\simeq (1-3) \times 10^{11}\ L_\odot$ (corresponding to rates of star formation of $SFR \sim 10-50\ M_\odot/yr$ if dominated by stellar activity, from Kennicut's calibration, see below).

At redshifts higher than $\sim 0.7$, this population appears to stop its evolution and then even reverse it. In parallel, a more luminous class of galaxies comes on stage. While providing a negligible contribution to the LF's at low-z, this very luminous and fast-evolving population overcome the LIRGs at $z>1$. At their peak activity epoch around $z\sim 1.5$ to 2.5, they show bolometric luminosities in the ULIRG range ($L_{bol}\simeq 10^{12}\ L_\odot$, SFRs of several hundred $M_\odot/yr$ under the same assumptions as above).  

Altogether, the IR sources manifest a marked bymodal behaviour, particularly evident at $z>1.5$. At these redshifts, although the direct estimates of the LF's available to us from Figs. \ref{LF15} and \ref{LF24} do not probe deeply in luminosity and are based on moderately numerous samples of a few hundreds sources (Rodighiero et al. 2010), the overall shape of the LFs is constrained by our integrated data. Still with reference to the two highest-z panels in Fig. \ref{LFbol}, on one hand, the pronounced ULIRG peak is required to explain the high-z tail in the redshift distributions of Figs. \ref{dz24} and \ref{dz850}, the high mm number counts in Figs. \ref{mm}, \ref{blast}, and the observed bright end of the LFs themselves. On the other hand, the low-level normalization of the LIRG population contribution at lower luminosities and the still flat shape of the LFs at these redshifts are constrained by the observed convergence of the 24 and 70 $\mu$m counts and by the requirement not to exceed the COBE background. 

At $z>1.5$ the bulk of the activity appears to happen in very luminous objects: more than 2/3 of it occurs in galaxies more luminous than $2\ 10^{11}\ L_\odot$. These sites of active SF and obscured accretion are presumably also already massive, in consideration of the roughly linear proportionality between stellar mass and bolometric IR emissivity found by various authors (Perez-Gonzales et al. 2005; Caputi et al. 2006; Daddi et al. 2007).

As illustrated in Fig. \ref{dz850}, our modelling of the statistical properties of this luminous population fits those observed for the SCUBA galaxies, in terms of both the detailed number counts and redshift distributions. 
Current interpretation identifies these ULIRG objects, and the SCUBA sources with the progenitors of the massive spheroidal galaxies, where the conversion of large amounts of molecular gas into stars are considered to explain the huge IR luminosities (e.g. Franceschini et al. 1994; Lilly et al. 1999; Blain et al. 2002).
Our current modelling of the high-z source statistics predicts that only a small fraction of the SCUBA objects should be found at $z>3$ (see also Rowan-Robinson 2009).

Let us now make a working assumption that the bulk of the energetics associated with the IR source populations (except of course for the few type-I AGNs described in our model) originates from young stellar activity. If so, the evolutionary patterns in Figs. \ref{rho}, \ref{LFbol}, and \ref{Ldens} would be interpreted to a large extent as characters in the evolutionary history of the formation of luminous (massive) stars in the universe (massive stars are largely responsible for the heating of dust in star-forming regions and for dust heating in luminous IR galaxies).     We test now this assumption with the existing complementary information on the star-formation history in galaxies based on observations of the integrated stellar mass densities.

To this end, we need to convert our best-fit model distribution with redshift of the comoving bolometric IR emissivity in Fig. \ref{rho} into a comoving rate of stellar formation, in such a way that data in this figure can be read in terms of a rate of star formation. 
This conversion is a sensible function of the initial mass function (IMF) ruling star-formation. The IMF determines not only the relationship between $L_{IR}[8-1000 \mu m]$ and the $SFR$, but also the mass fraction $f_{\ast}$ of stars remaining at a given galactic epoch, considering the fraction of baryonic gas ejected from stars during the evolution and returned to the ISM, or freezed into collapsed objects, black-holes, white-dwarfs, or neutron stars. 
The fraction $f_{\ast}$ is obviously a function of the time elapsed from formation to that of the observation, and depends on the stellar IMF (Madau and Pozzetti 2000; Cole et al. 2001).

To keep our analysis as general as possible, we have made use of a set of different IMFs, including significant variations in the Salpeter IMF, by increasing or decreasing the relative incidence of high-mass or low-mass stars.
For example, an IMF that is more top-heavy than the standard Salpeter one reduces both the rate of SF for a given observed $L_{IR}$ and the average stellar residual fraction $f_{\ast}$ at any galactic epoch. The opposite is obtained by reducing the relative fraction of high-mass stars (like in the Scalo [1986] IMF).
To derive the relationship between stellar mass and bolometric luminosity, we used the \textit{Pegase} model (Fioc and Rocca-Volmerange 1997) by assuming constant SFR during 100 Myrs and \textit{Category-1} stellar isochrones (Bressan et al. 1993). 
We also assumed that the whole UV flux from young stars is absorbed by dust and re-radiated in the IR, a situation  appropriate for LIRGs and ULIRGs (while it is less clear for normal spirals, which do not, however, contribute significantly to our energetics).
For the standard Salpeter's (1955) power-law form for the stellar initial mass function, $dn/d \log M \propto M^{-m}$ for stars between 0.1 and 120 $M_\odot$, the calibration between the bolometric luminosity $L_{IR}[8-1000 \mu m]$ and the $SFR$ is
\begin{equation}
SFR\ [M_\odot/yr] = K \ L_{IR}[8-1000 \mu m]/L_\odot ,
\label{ken}
\end{equation}
with
\[  K  \simeq 1.63\ 10^{-10} , \]
consistent with Kennicutt's (1998) reported value. 
We have also considered IMF power-law functions with different slopes $m$, in particular flatter, \textit{top-heavier} IMF's, for which the constant in the relation \ref{ken} can change quite significantly. Calibration values corresponding to different IMF's are reported in the last row of Table \ref{table0}, together with the stellar residual fractions $f_{\ast}(t)$ as a function of the stellar population age. Data for the Salpeter mass function, mass functions of power-law shape between 0.1 and 120 $M_\odot$ ($dn/d \log M \propto M^{-m}$) with larger fractions of massive stars (Top-heavy 1, 2, 3, corresponding to $m=1.15, 1, 0.85$), as well as for the Scalo (1986) and the Kroupa et al. (1993) IMFs, appear in the table.
Going from the Scalo IMF to the top-heavier ones, the stellar residual fraction $f_{\ast}(t)$ and the $SFR$ calibration decrease systematically.

From Eq. \ref{ken}, a comoving stellar mass density produced by these IR star-forming galaxies (type-I AGNs excluded) can be easily computed as
\begin{equation}
\rho_{star}(>z) = \int_z^{z_{max}}  dz' \rho_{SFR}(z')\cdot \left(\frac{dt}{dz'}\right)
\cdot f_\ast [t(z)-t(z')] ,
\label{rhostar}
\end{equation}
where $\rho_{star}(>z)$ is the integrated comoving stellar mass function at redshift $z$, $\rho_{SFR}$ the comoving star-formation rate density, scaled from $\rho_{IR}$ (e.g. as in Fig. \ref{rho}) from eq.\ref{ken}, and where 
\begin{equation}
\frac{dt}{dz} = \frac{1}{H_0\cdot (1+z)\cdot \sqrt{(1+z)^2 (1+\Omega_m z)-z(2+z)\Omega_\Lambda )}  }
\label{dtdz}
\end{equation}
is the differential of cosmic time as a function of redshift.  The function $f_\ast[t(z)-t(z')]$ corrects from the mass converted into stars at redshift $z'$ to the present stellar mass at the time $t(z)$. The $f_\ast$ values in Table \ref{table0} account also for the fraction of baryonic gas condensed into collapsed objects.

\subsection{Constraining the history of star formation and AGN accretion}
\label{SFRH}

\begin{figure}[!ht]
\centering
\includegraphics[angle=0,width=0.5\textwidth,height=0.45\textwidth]{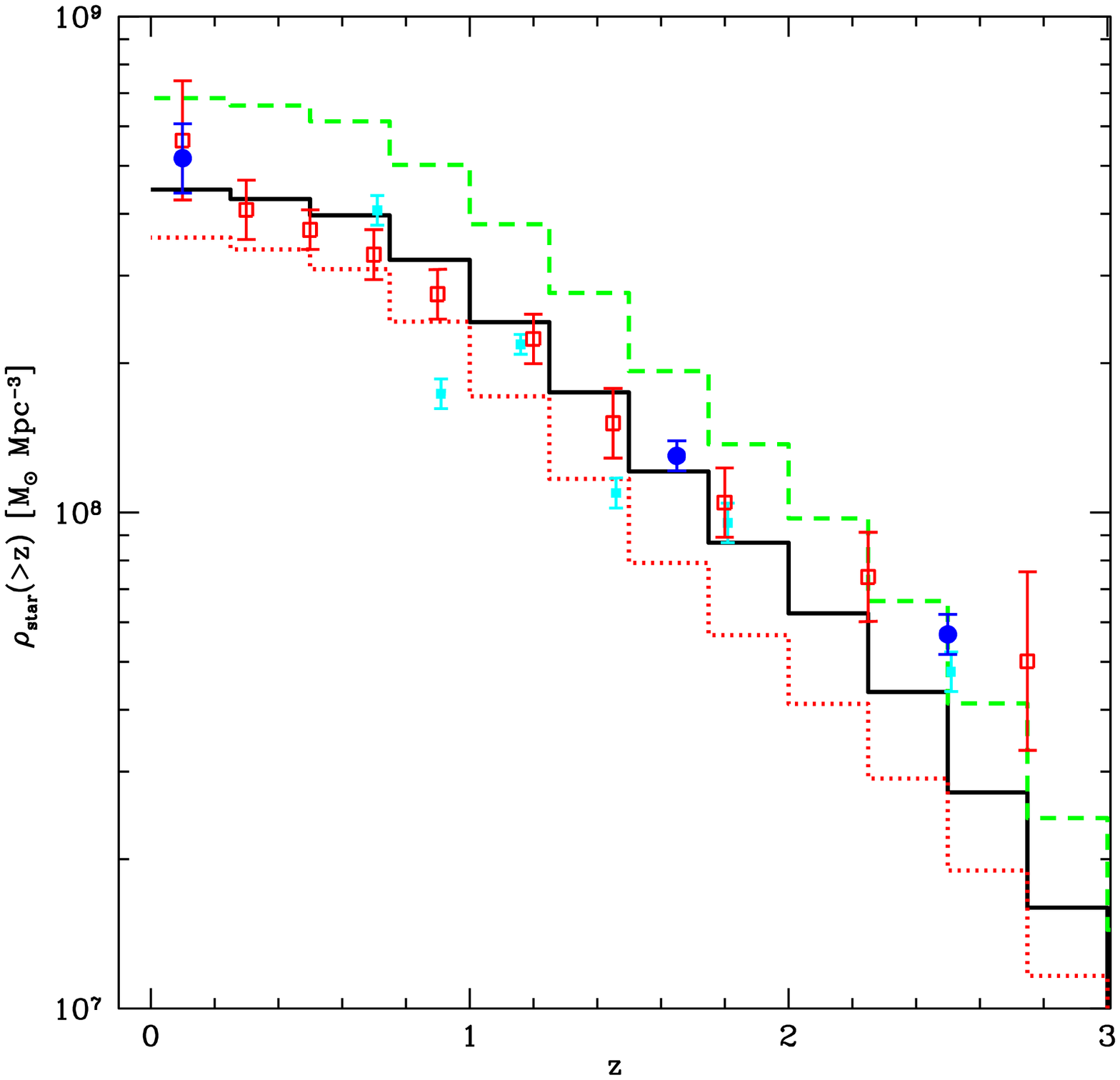}  
\caption{ 
Predictions of the cumulative distribution of the comoving stellar mass density as a function of redshift. Units are solar masses per comoving cubic Mpc.  
The black continuous-line histogram corresponds to our adopted model for the SF history and a standard Salpeter IMF, the red dotted-line histogram to a \textit{Top-heavy 1} IMF, the green dashed line to a Kroupa et al. IMF (see text for details).  
The red squared datapoints are estimates of the integrated stellar mass density by Perez-Gonzalez et al. (2008) based on deep Spitzer/IRAC surveys, while the cyan data are from Fontana et al. (2006; as indicated by the authors, we applied a 25\% correction to these data to make their mass determination consistent with others in the literature). Blue filled circles are from Marchesini et al. (2009).
}
\label{rhom1}
\end{figure}

We report in Fig. \ref{rhom1} predictions by the best-fit IR-evolution model concerning the integrated stellar mass already assembled by the various populations as a function of redshift and compare them to the observed integrated stellar mass densities. The three predictions correspond to three different adopted IMFs: the standard Salpeter ($m=1.35$), the \textit{Top-heavy 1}, and the Kroupa et al. (1993) IMFs.
Following our analysis in Sect. \ref{origin}, in this calculation we have assumed that 80\% of the IR emissivity by LIRGs, and 70\% of that by ULIRGs are due to star formation, with the rest of the radiant energy coming from an obscured AGN. This corresponds to adopting an AGN fractional contribution to the luminosity density of $f_{AGN}=0.3$ for the ULIRG sources.

\begin{figure*}[!ht]
\centering
\includegraphics[angle=0,width=0.8\textwidth,height=0.7\textwidth]{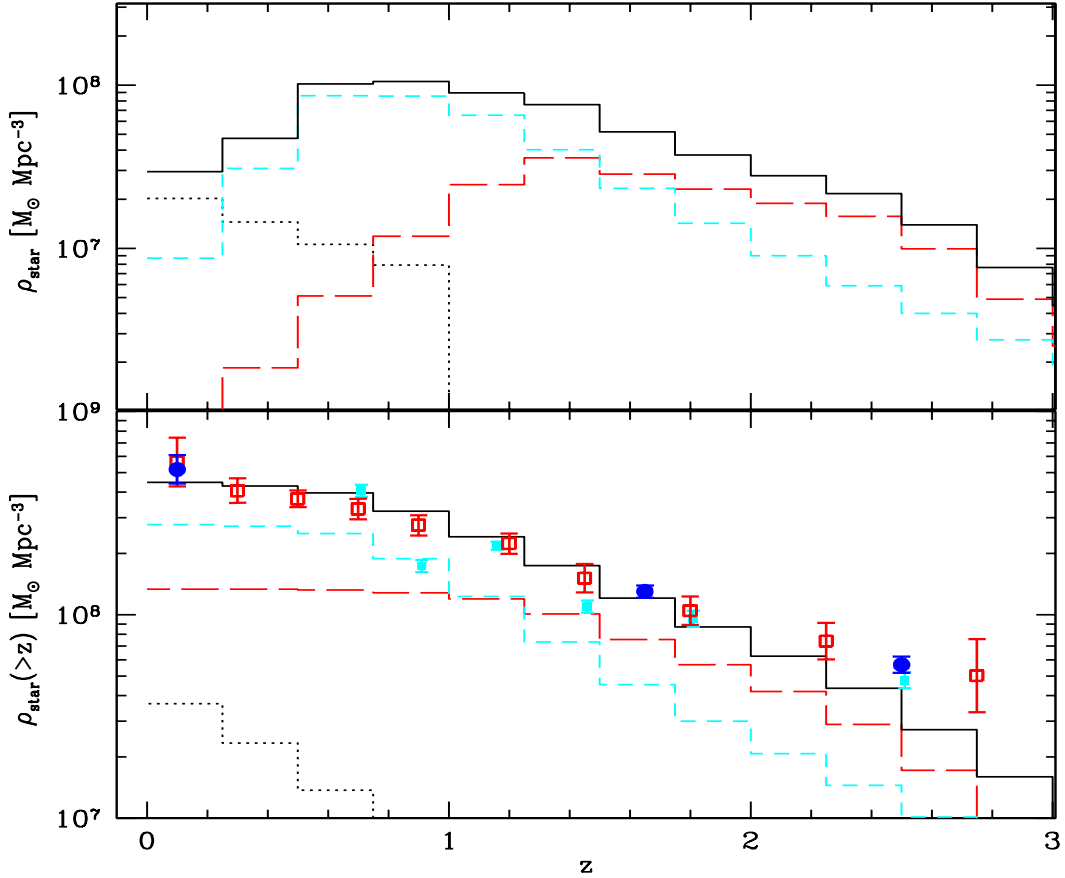}  
\caption{The stellar mass produced by IR starburst galaxies, according to our reference model, in various redshift bins. Units are solar masses per comoving cubic Mpc.  Line types as in Fig. \ref{rho}.
 \textit{Lower panel:} the cumulative distribution of stellar mass as a function of redshift. The red squared datapoints are estimates of the integrated stellar mass density by Perez-Gonsalez et al. (2008) based on deep Spitzer/IRAC surveys, while the cyan data are from Fontana et al. (2006; as indicated by the authors, we applied a 25\% correction to these data to make their mass determination consistent with others in the literature). Blue filled circles are from Marchesini et al. (2009).
The black line histogram corresponds to our adopted best-fit model for the SF history (in cyan short-dashed and red long-dashed those corresponding to our LIRGs and ULIRGs).
\textit{Upper panel: } instantaneous stellar mass produced per redshift bin. The normalization factors adopted here are 80\% of emitted energy by stars in LIRGs and 70\% by ULIRGs (no correction for gas recycling by stars).
}
\label{rhom}
\end{figure*}

These predictions for the cumulative stellar mass density are compared with data derived by Perez-Gonzalez et al. (2008), Fontana et al. (2006), and Marchesini et al. (2009) based on a large sample of galaxies selected from deep Spitzer/IRAC surveys from 3.6 to 8 $\mu$m. The IRAC data are essential to achieving good accuracy in the estimate of the stellar mass content in high-redshift galaxies (see also Bundy et al. 2006; Franceschini et al. 2006; Fontana et al. 2006, among others). 
We have neglected here the (small) correction factors to the observed cumulative stellar mass densities in Fig. \ref{rhom1} corresponding to a change in IMF, because the $M/L$ ratio used for the stellar mass estimate based on IRAC fluxes is weakly dependent on $m$.

As we can see, the three adopted IMFs correspond to significantly different predictions for the integrated stellar mass in galaxies as a function of redshift \footnote{Of course, changing the two parameters $f_{AGN}$ and the average stellar residual fraction $f_{\ast}$ does not alter our best-fit modelling of the evolutionary IR emissivity in any way (e.g. in Fig. \ref{rho}), but only our interpretation of the primary power sources producing it.}. A best fit to the data is clearly achieved with the Salpeter IMF, while the Kroupa and \textit{Top-heavy 1} IMFs appear inconsistent with the data, producing systematically higher and lower stellar mass distributions as a function of redshift.
Stronger departures from the Salpeter's function, like the Scalo (1986) and the top-heavy IMFs 2 and 3 in Table \ref{table0}, bring into even more serious conflicts with the data.

There are essentially two sources of uncertainty in our analysis. The first one is related with the currently still uncertain evolution of the bolometric luminosity density at $z>2.5$, see Sect. \ref{EVOL}: a higher emissivity at these redshifts, still partially consistent with data in our hand, would increase the stellar mass density at high-z, and consequently  also slightly rise the cumulative density at lower-z.

The second uncertainty comes from the poorly known contribution of AGN gravitational accretion to the observed IR bolometric emissivity at high redshifts. The AGN fractions assumed here are the closest to the data that we discussed in Sect. \ref{origin}. However, a precise estimate of the contribution of a nuclear non-thermal power source to the bolometric light is still not achievable with current data, particularly for high-redshift objects (while future studies by \textit{Herschel}, JWST, and ALMA, among others, will offer further means to disentangle stellar and AGN contributions during the evolution of cosmological sources).


In any case, a moderately \textit{Top-heavy 1} IMF, like that corresponding to the dotted red line in Fig. \ref{rhom1}, cannot be made entirely consistent with the data in the figure by decreasing the AGN fraction, because our currently assumed value is already close to the minimum (it is now $<30\%$). Perhaps this slightly top-heavy IMF could be marginally reconciled with the data if we consider the additional contribution - not accounted for in our analysis - of star formation in UV-selected galaxies at high redshifts. The contribution of the latter to the bolometric light, however, is estimated to be only of the order of 20\% of the bolometric emission (e.g. Hopkins and Beacon 2006; see also Madau et al. 1996).

Also, our prediction based on the Kroupa IMF is difficult to reconcile with the data, because it would require adopting a very large AGN fraction even for moderate- and low-luminosity sources, not justified by the present data. Indeed, the detailed shape of the dashed line in Fig. \ref{rhom1}, with a larger excess for the stellar density at lower-z, would require a higher AGN fraction in lower luminosity galaxies, contrary to any expectations.
Note, finally, that the inconsistency of the dashed-line solution with the data would remain large even when adopting a Salpeter IMF for the LIRG and using a Kroupa IMF for the ULIRGs.

Altogether, our comparison of the source bolometric luminosity density with the observed stellar mass density as a function of redshift reveals that these data are entirely consistent with the adoption of a universal Salpeter-like IMF, and contributions by AGN accretion to the IR energetics around 20 to 30\%.
It is interesting to stress that this result appears somewhat at variance with earlier findings (e.g. Hopkins and Beacon 2006;  Perez-Gonzalez et al. 2008). These authors have indicated a possible redshift- (or luminosity-) dependence of the average stellar IMF to reconcile the observational data on the stellar mass and luminosity densities. According to them, a top-heavy IMF would be required at $z>1.5$ to explain their very high estimated bolometric source emissivity at those redshifts.
The lower high-z emissivity indicated by our analysis can be made consistent more easily with a universal (Salpeter-like) IMF and standard AGN fractions.

In Fig. \ref{rhom} we provide further details about our best-guess fit to the comoving stellar mass density distribution, with the separate LIRG and ULIRG contributions (lower panel), and the instantaneous stellar mass produced per redshift bin with no correction for gas recycling by stars. As expected, the most luminous ULIRG population dominates the stellar production at $z>1$, while the lower luminosity starbursts do it at lower redshifts. Non-evolving spirals contribute significantly only at $z<0.5$ compared to the other populations.

A substantial AGN fraction among luminous IR MIPS sources at high redshifts has been reported by many authors, based on hard X-ray observations, population stacking X-ray measurements and and IRS spectroscopy (Yan et al. 2006; Sajina et al. 2007; Daddi et al. 2007; Polletta et al. 2008; Martinez-Sansigre et al. 2008, Watabe et al. 2009; Fiore et al. 2008, 2009, among many others). X-ray stacking by Fiore et al. shows that as many as 67\% of MIPS objects with $S_{24}>550\ \mu$Jy and $z>0.6$ contain an AGN, many of which are Compton-thick. Values consistent with this are inferred by Watabe et al. (2009) from IRS mid-IR spectroscopy. These are certainly to be considered as upper limits to the fraction of high-z galaxies containing trace AGN emission.
In addition, none of these studies can reliably constrain the fractional contribution of the AGN to the bolometric source emission. Our completely independent analysis suggests that, although a substantial fraction high-z IR sources might contain an AGN (e.g. Daddi et al. 2007), the AGN contribution (from gravitational accretion) to the bolometric flux should be small on statistical grounds ($<30\%$).

Local remnants of the high-redshift activity revealed by IR data would be expected to be massive spheroidal galaxies and super-massive black-holes in their centers. The stellar component would make around 30\% of the local stellar mass density in the universe on average, but would be expected to concentrate on high-mass systems and high-density environment, as a consequence of their highly biased formation at high redshifts. Preliminary angular correlation analyses support this view (Magliocchetti et al. 2006), but more extensive surveys with \textit{Herschel} will be needed to prove it.

If we consider, in addition, the close similarity of the evolutionary patterns as a function of source luminosity described in Fig. \ref{Ldens} for the IR-luminous sources with those emerged for the type-I X-ray AGNs in Hasinger, Miyaji, and Schmidt (2005) and Bongiorno et al. (2008), all this strongly supports the view of a tight link between SF and AGN activity in cosmic sources. 
Only high-resolution investigations with future instrumentation will reveal whether this is only a statistical connection or the signature of a deeper physical relationship among these forming structures, such as a ruling effect of one on the other (e.g. AGN feedback on the circum nuclear SF).

As a word of caution, our results from the comparison of the stellar mass and SFR densities are based on a combination of mid-IR (mostly 24 and 70 $\mu$m) data on the redshift-dependent luminosity functions, together with far-IR and sub-mm measurements of integrated quantities, like number counts and observations of the CIRB background radiation (spectral intensity and cell-to-cell fluctuations).
Although the amount of current data strongly constrain our model solutions, we still lack a direct evaluation of the bolometric luminosity functions and IR emissivity of cosmic sources as a function of the cosmic time. The latter will be attempted soon on the basis of the faint far-IR and sub-mm source samples that the \textit{Herschel Space Observatory} is starting to produce.
\textit{Herschel} will provide us with a unique opportunity to largely resolve some remaining ambiguities about the history of energy production by cosmic sources and its interpretation.  Data from its sensitive photometric imaging cameras, PACS and SPIRE, cover a wide wavelength interval between 70 and 500 $\mu$m, where the bulk of the CIRB energy resides.
The large systematic programme of cosmological surveys planned with \textit{Herschel} over the whole far-IR and sub-mm spectral region (e.g. Griffin et al. 2007; 2010) will soon both resolve the CIRB background into its individual sources and, in combination with pre-existing optical and \textit{Spitzer} data over the survey areas, allow us to extensively sample the multi-wavelength luminosity functions.  X-ray and radio data will offer additional diagnostic tools to disentangle the contributions of the two energy production mechanisms.


\section{Conclusions}
\label{conclusion}

We report on our analysis of a large IR database on high-redshift galaxies at long wavelengths, the bulk of which we obtained from recently completed surveys with the \textit{Spitzer Space Telescope}, including unpublished results. 
\textit{Spitzer} data are complemented by sub-mm surveys from the balloon experiment BLAST, preliminary results of the \textit{Herschel} observatory, millimetric observations with large ground-based telescopes, as well as diffuse extragalactic background measurements, mostly from COBE. 

With these data in hand, we have substatially updated our previous (AF2001) modelling of galaxy evolution. Our new results agree with AF2001 at $z<1$, but much improve their analysis at redshifts $>1$, where a very luminous source population dominating the IR activity has been established by the new data.

Our main results are summarized in the following.

   \begin{enumerate}

\item
The current data completely confirm earlier indications of a very rapid increase in galaxy long-wavelength volume emissivity with redshift up to $z\simeq 1$, paralleled by an increased incidence of dust extinction and thermal dust reprocessing in high-redshift sources, with respect to locally observed galaxies. 
This is the fastest evolution rate observed for galaxies at any wavelengths [$\rho(z)\propto (1+z)^{4}$ if averaged over the whole galaxy population].
Confusion-limited number counts at longer wavelengths (from 70 to 1100 $\mu$m) completely support these findings.

\item
All the present data require that the fast evolution observed from z=0 to 1 flattens around redshift 1 and keeps approximately flat, at least up to $z\simeq 2.5$.  At higher redshifts, the comoving emissivity may be required to decrease to avoid exceeding the COBE observed intensity of the CIRB, but more data from \textit{Herschel} will be needed for more definite conclusions. 
In any case, there is a clear tendency in our redshift-dependent bolometric comoving energy density (Fig. \ref{rho}) to stay lower than various published results, particularly at higher redshifts than $z\sim 1.5$, based on either large extinction corrections, or large spectral extrapolations.

\item
Our exploration of the high-redshift ($z>1$) universe in the far-IR and sub-mm has provided evidence of a population of very luminous galaxies becoming dominant at $z>1$. The comoving emissivity of the latter is maximal at $z\simeq 2$ and shows a much faster convergence with cosmic time than lower luminosity systems, whose maximal activity is set around $z\simeq 1$. Then an earlier phase of formation for the most luminous/massive galaxies/AGNs is indicated and provides supporting evidence of the peculiar evolutionary pattern named \textit{downsizing}. At the highest redshifts probed by the present analysis ($1.5<z<3$), the luminosity functions of galaxies appear to be dominated by ultra-luminous sources ($L_{bol}\simeq 10^{12}\ L_\odot$), while the comoving number density of lower luminosity galaxies is low and their LFs flat (consistent with being not steeper than in the local universe).
These flat shapes of the LFs at any redshifts indicated by the present analysis (Fig. \ref{LFbol}) might require significant tuning or modifications of galaxy formation models.

\item
The IR emissions by cosmic sources analysed here include the contributions of both massive-star formation and dust-obscured AGN accretion.  Many attempts have been made to disentangle the two, mostly based on deep hard X-ray imaging and \textit{Spitzer} IRS spectroscopy.
All these analyses, however, remain essentially inconclusive in so far that, in high-column-density media, the hard X-ray flux is completely removed by Compton scattering and the mid-IR hot-dust AGN emission may become optically thick.
To gain more insight, we followed a complementary road of comparing our results on the comoving IR emissivity of sources with recent estimates of the redshift-dependent stellar mass functions of galaxies based on deep (Spitzer/IRAC) near-IR surveys. In our analysis, the comparison of the istantaneous SFR with the integrated mass assembled in stars relies on two model parameters: the AGN fractional contribution to the bolometric IR emission, $f_{AGN}$, and the slope $m$ of the stellar IMF.
Unlike photonic remnants like the background radiation, suffering a $(1+z)^{-1}$ redshift penalty factor, the stellar mass density is an unbiased tracer of the past SF activity at the high redshifts.

\item
Using our \textit{best-fit} evolutionary model, we find that the observed redshift-dependent bolometric source emissivity $\rho_{IR}$ can be naturally reconciled with the observed galaxy mass functions by adopting a universal Salpeter IMF ($m=1.35$) and standard AGN fractions $f_{AGN}\sim 20-30\%$ in LIRGs and ULIRGs at all redshifts. Previous analyses referred to higher estimated $\rho_{IR}(z)$ at $z>1.5$, hence suggested top-heavy IMFs for the more luminous galaxies at higher redshifts or higher AGN fractions.

\item
We finally caution that with current data we cannot probe deeply into the luminosity functions of high-redshift far-IR sources.
Statistical constraints were included in our combined dataset, in particular upper limits on LFs at the faint ends from the use of number counts and  the CIRB intensity (including those from cell-to-cell fluctuations). Average spectral shapes for populations of sources were derived based on the multi-wavelength information from the mid-IR to the millimetre.
Cosmological surveys with the \textit{Herschel} observatory will soon provide us with data to resolve the present residual uncertainties about the history of energy production by cosmic sources and its interpretation.

   \end{enumerate}

\begin{acknowledgements}
This work was supported by the Italian Space Agency (ASI) under contracts I/005/07/0 (Herschel Fase-E) and I/057/08/0 (SPICA-SAFARI) and research funding by the University of Padova.  
We warmly thank an anonymous referee for useful comments.
\end{acknowledgements}


\begin{appendix} 
\label{A}

\section{Contribution of cosmic sources to the background radiation: the formalism}

A simple formalism relates background intensity and cell-to-cell anisotropies to the statistical properties (luminosity functions and number counts) of the contributing sources.

\subsection{Source contribution to the background intensity}

The differential number counts (sources/unit flux interval/unit solid angle) at a given flux 
$S$ write as
\begin{equation}
{dN \over dS} = 
\int_{z_l}^{z_h}\,dz\,{dV\over dz}\,{d \log L(S;z)\over dS}\, 
\frac{dN}{dL}[L(S,z),z] \label{dNdS}
\end{equation}
where $\frac{dN}{dL}[L(S,z),z]$ is the epoch-dependent (comoving) luminosity function (see e.g. Figs. \ref{LF15}, \ref{LF24}, and \ref{LF35}) and $dV/dz$ is the differential (comoving) volume element.
Flux $S$ and rest-frame luminosity $L$ are related by 
\begin{equation}
S_{\Delta \nu} = {L_{\Delta\nu} K(L,z) \over 4\pi d_L^2}, \label{S}
\end{equation}
where $d_L$ is the luminosity distance and
$K(L,z)= (1+z) {L[\nu (1+z)]\over L(\nu)} $ the K-correction.  
The contribution of unresolved sources (sources fainter than the detection limit $S_d[\nu_0]$ at the frequency $\nu_0$) to the background intensity is given by
   \begin{equation}
   \begin{array}{l}
I(<S_d[\nu_0]) = \frac{1}{4\pi} \int_0^{S_d[\nu_0]}{dN\over dS} S\, dS =
\\ \\
= \frac{1}{4\pi} \frac{c}{H_0} \int_{0}^{z_{max}} dz\ \frac{j[\nu_0(1+z),z]}{(1+z)\times[(1+z)^3\Omega_m + \Omega_\Lambda]^{1/2}} ,
   \end{array}
   \label{4}
   \end{equation}
for the rather general case of a flat universe with $\Omega_m + \Omega_\Lambda=1$. Here, $j[\nu_0(1+z),z]$ is the redshift-dependent galaxy comoving volume emissivity:
%
%
\begin{equation}
j[\nu_0(1+z),z] = \int_{L_{\rm min}}^{\min\left[ L_{\rm max}, L(S_d,z) 
\right] } d\log L 
\ L \ \frac{dN}{dL}(L,z) K(L,z), 
\label{eq:22} 
\end{equation}
where $L_{\rm min}$ and $L_{\rm max}$ are the minimum and the maximum source
luminosities. From Eq.(A.3) we can see that, when the counts converge
like $dN/dS \propto S^{-2}$ or flatter, the contribution by faint sources to the
background intensity almost becomes insensitive to the source minimum flux 
[$I\propto ln(S_{min})$ or less]. This property has been used by Madau \& Pozzetti (2000)
to estimate the optical background intensity from ultra-deep HST
counts of galaxies, by exploiting the convergence of the optical counts fainter than
$m_{AB}\sim 22$. Similar properties of convergence of the faint IR source counts have been used by Franceschini et al. (2008) to estimate the contribution of IR galaxies to the CIRB.

%

\begin{figure*}[!ht]
\centering
\includegraphics[angle=0,width=0.95\textwidth,height=0.75\textwidth]{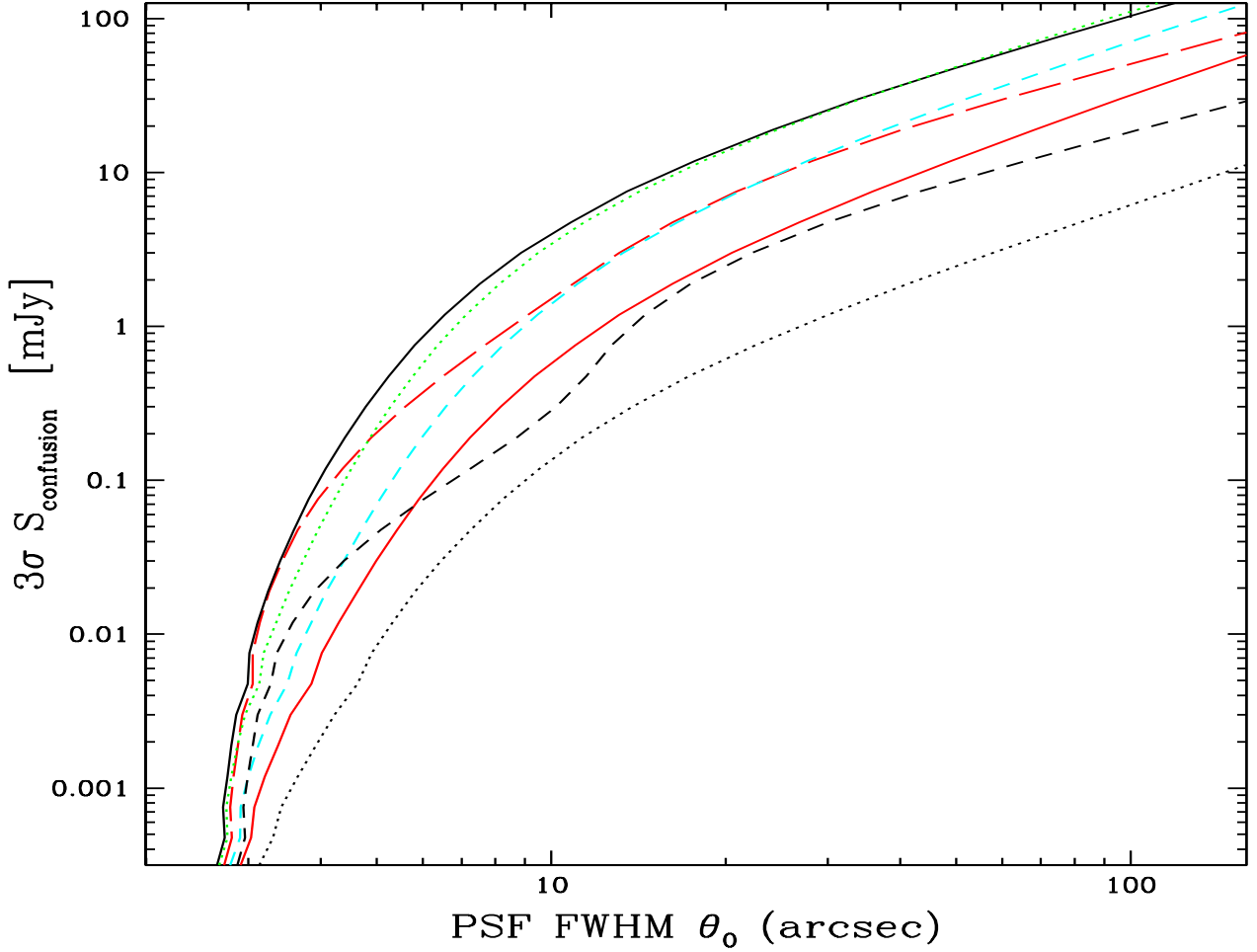}  
\caption{Estimated 3$\sigma$ confusion limits as a function of the FWHM angle $\theta_0$ of a Gaussian fit to the PSF (see Appendix A and B), $q=3$, for observations at various effective wavelengths. The latter are reported with the following colour and line codes:
black dots: $\lambda=$35 $\mu$m;  
red continuous: 70 $\mu$m;  
cyan short dash: 100 $\mu$m;  
green dots: 160 $\mu$m;  
black continuous: 250 $\mu$m;  
red long dash: 500 $\mu$m;  
black short dash: 850 $\mu$m.
}
\label{thetaconf}
\end{figure*}

\subsection{Small scale intensity fluctuations}

In addition to the average integrated flux by all sources in a sky area, the background 
radiation also contains spatial information, the cell-to-cell fluctuations, which can 
be used to further constrain the source flux distribution and spatial correlation properties.
The usually most important contribution to the cell-to-cell intensity fluctuations 
comes from the stochastic nature of the spatial distribution of sources among
elementary cells with an effective solid angle $\omega _{\rm eff}$ (see below). They can be expressed as
\begin{equation}
(\delta I)^2 = {\omega_{\rm eff}\over 4\pi}\int_0^{S_d} 
S^2\,{dN \over dS}\, dS .
\end{equation}
What is really measured, however, is not 
the flux $S$ but the detector's response $x=f(\vartheta , \varphi)$, 
$f(\vartheta , \varphi )$ being the angular power pattern of the detector. 
Let 
\begin{equation}
R(x) = \int_0^\infty dN\left[x/f(\vartheta , \varphi )\right]/ dS \cdot d\omega/f(\vartheta ,\varphi ) 
\end{equation}
be the {\it mean number of source responses of intensity $x$ in the beam}.
For a Poisson distribution of the number of sources producing a response $x$, its variance equals the mean $R(x)dx$. 
Adding the variances of all responses up to the cutoff value $x_c$ (brighter sources are considered to be individually detected) gives the contribution of unresolved sources to fluctuations:
\begin{equation}
(\delta I)^2 = \int_0^{x_c}x^2\,R(x)\,dx. 
\label{eq:47}
\end{equation}
The cutoff $x_c$ is chosen to be some factor $q$ times $(\delta I)^2$; usually $q= 3$--5. The rms background fluctuations ($\delta I$) imply a sky noise $\sigma_{conf}={\langle(\delta I)^2\rangle}^{1/2}$ for observations with spatial resolution $\omega_{eff}$. 
Assuming a Gaussian fit with $FWHM=\theta_0$ to the image's PSF:
\begin{equation}
f(\theta)=exp[-4(\theta/\theta_0)^2 ln(2)]
\label{PSF}
\end{equation}
with $\theta_0\simeq 1.02\lambda/D$ to represent the Airy function for diffraction-limited observations, we have 
\begin{equation}
\sigma^2 = (\delta I)^2 = \pi (\theta/\theta_0)^2 \ \Sigma(x_c) , 
\label{eq:47}
\end{equation}
\begin{equation} \Sigma(x_c)=\int_0^{x_c}dx x^2 R(x)= 
\label{Sigma}
\end{equation}
$$ = \int_0^{x_c} dx x^2 \int_0^\infty d\psi N[x/f(\psi)]exp(4\psi ln(2))$$
with $\psi=\theta/\theta_0$.  In practice, it is custom to compute the confusion limit $\sigma_{conf}$ from the relation
\begin{equation}
\theta_0=\sigma / \sqrt{\pi \Sigma(x_c)}
\label{recursive}
\end{equation}
by assuming $\sigma=q x_c$ and $q=3$ or 5 (e.g. Franceschini et al. 1989).  The confusion limit (at 3 or 5 times the rms confusion noise $\sigma_{conf}=\sigma$) is then easily computed with the zero's of relation \ref{recursive}.

The integrated signal $D$ recorded by the detector is the sum of the responses $x$ due to all sources in the angular resolution element.    Its probability distribution function $P(D)$ is informative on the amplitude and slope of counts of unresolved sources.
Scheuer (1957) has shown that, under the assumption of a random sky distribution of sources, its Fourier transform, $p(\omega)$, is a simple function of the FT $r(\omega)$ of $R(x)$:  $p(\omega) = \exp [r(\omega) - r(0)]$.  It follows:
\begin{eqnarray}
P(D)  =  \int _{-\infty}^{\infty} p(\omega) \exp(-2\pi i \omega D)\,d\omega = \\
\int _{-\infty}^{\infty} \exp\left[r(\omega)-r(0)-2\pi i \omega D\right]
\,d\omega =        \\
2 \int _{0}^{\infty} \exp \left\{-\int _{0}^{\infty} R(x) 
\left[1-\cos(2\pi \omega x)\right] dx\right\}  \cdot 
\label{P(D)} 
\end{eqnarray}
$$
  \ \ \ \ \   \cdot \cos \left[ \int_{0}^{\infty} R(x) \sin(2\pi \omega x)\,dx - 
2\pi \omega D \right] \, d\omega. 
$$
This synthetic $P(D)$ has to be convolved with the (typically Gaussian) distribution of the instrumental noise to be compared with the observations. 
Rather than in terms of the total flux per sky beam $D$, it is more frequent to refer the fluctuation analysis to the flux per unit sky area $D_a=D/\omega_{eff}$, where $\omega_{eff}$ is the imager effective area and is defined as 
\begin{equation}
\omega_{eff}={\int_0^{x_c} x R(x) dx \over I(<x_c) }
\label{omega} 
\end{equation}
with $I(<x_c)$ being the residual background, from Eq. (\ref{4}). It turns out that
\begin{equation}
\omega_{eff} \simeq \int f(\vartheta) d\Omega = \pi \int_0^\infty d\theta^2 f(\theta) = \pi(\theta_0/2)^2/ln(2)
\label{sigma} 
\end{equation}
for our Gaussian representation of the PSF in eq. (\ref{PSF}).

Assumed that the number count distribution below the detection limit can be represented as a power-law, $N(>S)=K(S/S_k)^{-\beta}$, then eq. [\ref{P(D)}] can be integrated to get (Condon 1974):
\begin{eqnarray}
\sigma_{conf}= \left[{ q^{2-\beta} \over 2-\beta}\right]^{1/\beta} 
(\omega _{\sigma} \beta K)^{1/\beta}
S_k,   \\ 
\omega_{\sigma}=\int f(\vartheta ,\varphi)^\beta d\Omega
\label{sigma} 
\end{eqnarray}
which allows estimation of the slope of the counts ($\beta$) below the detection
limit from a given measured value of the cell-to-cell fluctuations $\sigma_{conf}$. 
This constraint on $N(S)$ applies down to a flux limit corresponding to 
$\sim 1$ source/beam. Assumed that $S_k$ represents the confusion limit ($S_k=q\times
\sigma_{conf}$) of a survey having an areal resolution $\omega_{eff}$, 
then eq. \ref{sigma} further simplifies to a relation
between the number of sources $K$ resolved by the survey (and brighter than
$S_k$) and the parameters $q$ and $\beta$:
\begin{equation}
K = {2-\beta \over \beta q^2} {1 \over \omega_{\sigma}}:
\label{conf}
\end{equation}
this implies the confusion limit to occur at the flux corresponding to an areal density 
of $(\beta q^2/[2-\beta])^{-1}$ sources per unit beam area $\omega_{\sigma}$. For Euclidean
counts and $q=3$, this corresponds to 1 source/27 beams. 

\end{appendix}

\begin{appendix} 
\label{B}
\section{Model predictions for future observations}

\subsection{Source number counts}

Predictions for (differential and integral) source number counts are reported in Tables \ref{table1} to \ref{table4}, based on our \textit{best-fit} model. 

In each table the left column reports the logarithm of flux, and the second and third columns give the (Euclidean-normalized) differential and integral total number counts for the first effective wavelength indicated in the table caption. The fourth and fifth columns give the differential and integral counts for the second wavelength in the caption.

The units for the differential counts are the number of sources per square degree per unit flux interval in mJy, divided by the Euclidean function $S^{-2.5}$. Those for the integral counts are number per square degree.

\subsection{Predictions for source confusion limitations}

We calculated predictions for the source confusion limits based on the \textit{best-fit} model for observations at various wavelengths using relation \ref{recursive} in Appendix A and calculating the zero's of the relation as a function of the FWHM $\theta_0$ of the observation at a given wavelength.    This definition of confusion is sometimes referred to as \textit{photometric confusion}, dealing with the photometric noise induced by source pile-up in the map's elementary beam. 

As mentioned in Appendix A, for a diffraction-limited imaging survey the FWHM angle $\theta_0$ of a Gaussian fit to the PSF Airy function is given by
\begin{equation}
\theta_0 \simeq \frac{1.02\ \lambda}{D}
\end{equation}
with $\lambda$ being the effective observational wavelength and D the primary collector diameter. If the imager is not diffraction-limited,  $\theta_0$ is in any case the PSF FWHM.
We report in Fig. \ref{thetaconf} our estimated 3$\sigma$ confusion limits for various far-IR and sub-mm wavelengths assuming a value $q=3$.

%
%
\begin{table}
\caption{Euclidean-normalized differential and integral number counts at $\lambda_{eff}=35$ and $70\ \mu$m by our best-fit model}             
\label{table1}      
\centering                          
\begin{tabular}{ c c c c c}        
\hline                 
log(S) & $log[dN/dS]$ & $log[N(>S)]$ & $log[dN/dS]$ & $log[N(>S)]$       \\
 $mJy$   & $deg^{-2} mJy^{-1}$ & $deg^{-2}$ & $deg^{-2} mJy^{-1}$ & $deg^{-2}$  \\    
\hline                        
      -1.9   &   1.718  &    4.916     &      1.759    &     5.059 \\
      -1.7   &    1.98  &    4.823	   &      2.014    &     4.991 \\
      -1.5   &   2.238  &    4.717	   &      2.269    &     4.919 \\
      -1.3   &   2.477  &    4.595	   &      2.526    &     4.843 \\
      -1.1   &     2.7  &    4.456	   &      2.783    &     4.759 \\
      -0.9   &   2.899  &    4.286	   &      3.038    &     4.665 \\
      -0.7   &   3.061  &    4.076	   &      3.287    &     4.557 \\
      -0.5   &   3.172  &    3.818	   &      3.523    &     4.432 \\
      -0.3   &    3.23  &    3.518	   &      3.738    &     4.286 \\
      -0.1   &   3.235  &    3.178	   &      3.926    &     4.113 \\
       0.1   &   3.202  &    2.816	   &      4.082    &      3.91 \\
       0.3   &   3.149  &    2.463	   &      4.202    &     3.676 \\
       0.5   &   3.097  &    2.122	   &      4.281    &      3.41 \\
       0.7   &   3.055  &    1.787	   &      4.323    &     3.113 \\
       0.9   &   3.027  &    1.492	   &      4.335    &     2.794 \\
       1.1   &   3.021  &    1.207	   &      4.324    &     2.467 \\
       1.3   &   3.021  &   0.8964	   &        4.3    &     2.138 \\
       1.5   &   3.025  &    0.601	   &      4.271    &     1.807 \\
       1.7   &   3.023  &   0.3178	   &       4.24    &     1.476 \\
       1.9   &   3.034  &  0.03021	   &       4.21    &     1.147 \\
       2.1   &   3.046  &  -0.2809	   &      4.179    &    0.8207 \\
       2.3   &   3.047  &  -0.5711	   &      4.154    &    0.4974 \\
       2.5   &   3.051  &  -0.8624	   &      4.131    &    0.1821 \\
       2.7   &   3.052  &   -1.176	   &      4.114    &   -0.1331 \\
       2.9   &   3.057  &   -1.481	   &      4.099    &   -0.4463 \\
       3.1   &   3.052  &   -1.781	   &      4.086    &   -0.7606 \\
       3.3   &   3.045  &   -2.091	   &      4.072    &    -1.079 \\
       3.5   &   3.042  &   -2.406	   &      4.056    &    -1.401 \\
       3.7   &   3.032  &    -2.72	   &      4.034    &     -1.73 \\
       3.9   &    3.01  &   -3.056	   &      4.006    &    -2.068 \\
\hline                                  
\end{tabular}
\end{table}
%

%
%
\begin{table}
\caption{Euclidean-normalized differential and integral number counts at $\lambda_{eff}=100$ and $160\ \mu$m by our best-fit model}             
\label{table2}      
\centering                          
\begin{tabular}{ c c c c c}        
\hline                 
log(S) & $log[dN/dS]$ & $log[N(>S)]$ & $log[dN/dS]$ & $log[N(>S)]$       \\
 $mJy$   & $deg^{-2} mJy^{-1}$ & $deg^{-2}$ & $deg^{-2} mJy^{-1}$ & $deg^{-2}$  \\    
\hline                        
     -1.9   &    1.811    &     5.171   &    1.872    &      5.307 \\
     -1.7   &    2.074    &     5.111   &    2.154    &      5.256 \\
     -1.5   &    2.335    &     5.047   &    2.429    &      5.202 \\
     -1.3   &    2.593    &     4.981   &    2.698    &      5.144 \\
     -1.1   &    2.853    &     4.911   &    2.965    &      5.081 \\
     -0.9   &    3.115    &     4.836   &     3.23    &      5.015 \\
     -0.7   &    3.375    &     4.752   &    3.494    &      4.945 \\
     -0.5   &    3.633    &     4.658   &    3.761    &      4.867 \\
     -0.3   &    3.882    &     4.549   &    4.027    &       4.78 \\
     -0.1   &    4.118    &     4.423   &    4.287    &      4.678 \\
      0.1   &    4.332    &     4.275   &    4.536    &      4.558 \\
      0.3   &     4.52    &     4.099   &    4.765    &      4.415 \\
      0.5   &    4.674    &     3.891   &    4.963    &       4.24 \\
      0.7   &    4.788    &     3.649   &    5.121    &      4.025 \\
      0.9   &    4.859    &     3.372   &    5.229    &      3.762 \\
      1.1   &    4.887    &     3.055   &    5.277    &       3.45 \\
      1.3   &    4.881    &     2.714   &    5.266    &      3.084 \\
      1.5   &    4.847    &     2.363   &    5.207    &      2.682 \\
      1.7   &    4.798    &     2.007   &    5.112    &      2.263 \\
      1.9   &    4.744    &     1.651   &    5.002    &      1.848 \\
      2.1   &    4.689    &     1.297   &    4.891    &      1.445 \\
      2.3   &    4.636    &    0.9495   &    4.786    &      1.051 \\
      2.5   &    4.587    &    0.6062   &    4.692    &     0.6677 \\
      2.7   &    4.544    &    0.2677   &     4.61    &     0.2964 \\
      2.9   &    4.504    &  -0.06816   &    4.541    &   -0.05632 \\
      3.1   &    4.466    &   -0.4033   &    4.484    &    -0.3998 \\
      3.3   &    4.431    &   -0.7385   &    4.439    &    -0.7369 \\
      3.5   &    4.397    &    -1.069   &    4.398    &     -1.074 \\
      3.7   &    4.367    &    -1.397   &    4.362    &     -1.412 \\
      3.9   &    4.342    &     -1.72   &    4.327    &     -1.748 \\
\hline                                  
\end{tabular}
\end{table}
%

%
%
\begin{table}
\caption{Euclidean-normalized differential and integral number counts at $\lambda_{eff}=250$ and $350\ \mu$m by our best-fit model}             
\label{table3}      
\centering                          
\begin{tabular}{ c c c c c}        
\hline                 
log(S) & $log[dN/dS]$ & $log[N(>S)]$ & $log[dN/dS]$ & $log[N(>S)]$       \\
 $mJy$   & $deg^{-2} mJy^{-1}$ & $deg^{-2}$ & $deg^{-2} mJy^{-1}$ & $deg^{-2}$  \\    
\hline                        
    -1.9  &    1.866   &      5.392  &   1.835    &    5.404  \\
    -1.7  &    2.188   &      5.348  &   2.188    &    5.364  \\
    -1.5  &    2.493   &      5.298  &   2.522    &    5.313  \\
    -1.3  &    2.785   &      5.242  &   2.834    &    5.253  \\
    -1.1  &    3.065   &      5.181  &   3.126    &    5.182  \\
    -0.9  &    3.339   &      5.113  &   3.401    &    5.102  \\
    -0.7  &     3.61   &      5.038  &   3.663    &    5.013  \\
    -0.5  &    3.876   &      4.956  &   3.912    &     4.91  \\
    -0.3  &    4.138   &      4.862  &   4.147    &    4.794  \\
    -0.1  &    4.392   &      4.754  &   4.369    &    4.663  \\
     0.1  &    4.631   &      4.624  &   4.576    &    4.514  \\
     0.3  &    4.847   &      4.469  &   4.761    &    4.344  \\
     0.5  &    5.032   &      4.282  &   4.917    &    4.139  \\
     0.7  &    5.172   &      4.054  &    5.04    &    3.896  \\
     0.9  &    5.261   &      3.774  &   5.118    &    3.614  \\
     1.1  &    5.297   &      3.442  &   5.139    &    3.279  \\
     1.3  &    5.267   &      3.065  &   5.085    &    2.873  \\
     1.5  &    5.166   &      2.625  &   4.931    &    2.363  \\
     1.7  &    5.009   &      2.112  &   4.696    &    1.742  \\
     1.9  &    4.822   &      1.575  &   4.418    &    1.093  \\
     2.1  &    4.626   &        1.1  &   4.144    &   0.5516  \\
     2.3  &    4.453   &     0.6731  &    3.92    &   0.1081  \\
     2.5  &    4.322   &     0.2747  &   3.769    &  -0.2791  \\
     2.7  &    4.224   &   -0.09666  &   3.672    &  -0.6411  \\
     2.9  &    4.149   &    -0.4531  &     3.6    &  -0.9954  \\
     3.1  &    4.088   &    -0.8021  &    3.54    &   -1.356  \\
     3.3  &    4.034   &     -1.151  &    3.48    &   -1.716  \\
     3.5  &    3.984   &     -1.503  &   3.424    &   -2.071  \\
     3.7  &    3.937   &     -1.849  &   3.375    &   -2.415  \\
     3.9  &    3.895   &     -2.188  &    3.33    &   -2.758  \\
\hline                                  
\end{tabular}
\end{table}
%

%
%
\begin{table}
\caption{Euclidean-normalized differential and integral number counts at $\lambda_{eff}=500$ and $850\ \mu$m by our best-fit model}             
\label{table4}      
\centering                          
\begin{tabular}{ c c c c c}        
\hline                 
log(S) & $log[dN/dS]$ & $log[N(>S)]$ & $log[dN/dS]$ & $log[N(>S)]$       \\
 $mJy$   & $deg^{-2} mJy^{-1}$ & $deg^{-2}$ & $deg^{-2} mJy^{-1}$ & $deg^{-2}$  \\    
\hline                        
   -1.9   &  1.859   &    5.382 &     2.049     &     5.214  \\
   -1.7   &  2.225   &    5.336 &	   2.346    &     5.101  \\
   -1.5   &  2.567   &    5.276 &	   2.585    &     4.955  \\
   -1.3   &  2.882   &      5.2 &	   2.765    &     4.775  \\
   -1.1   &   3.17   &    5.108 &	   2.877    &     4.584  \\
   -0.9   &  3.431   &    4.998 &	   2.932    &     4.396  \\
   -0.7   &   3.66   &    4.868 &	   3.006    &     4.251  \\
   -0.5   &  3.855   &    4.717 &	   3.157    &     4.164  \\
   -0.3   &  4.017   &    4.552 &	   3.376    &     4.097  \\
   -0.1   &  4.164   &    4.384 &	   3.644    &     3.993  \\
    0.1   &  4.322   &    4.223 &	   3.916    &     3.808  \\
    0.3   &  4.484   &    4.071 &	   4.105    &     3.552  \\
    0.5   &  4.642   &     3.88 &	   4.161    &     3.219  \\
    0.7   &  4.781   &    3.629 &	   4.064    &     2.759  \\
    0.9   &  4.846   &    3.312 &	   3.812    &     2.116  \\
    1.1   &  4.796   &    2.899 &	   3.425    &     1.271  \\
    1.3   &  4.607   &    2.338 &	   2.952    &    0.4295  \\
    1.5   &  4.301   &    1.607 &	   2.491    &   -0.3523  \\
    1.7   &  3.926   &   0.8353 &	   2.127    &   -0.9304  \\
    1.9   &  3.558   &   0.1836 &	   1.862    &    -1.363  \\
    2.1   &  3.266   &  -0.3064 &	   1.686    &    -1.769  \\
    2.3   &  3.076   &  -0.6918 &	   1.558    &    -2.198  \\
    2.5   &  2.961   &   -1.048 &	   1.443    &    -2.623  \\
    2.7   &  2.888   &   -1.411 &	   1.328    &     -3.04  \\
    2.9   &  2.818   &   -1.788 &	   1.207    &    -3.447  \\
    3.1   &  2.752   &   -2.159 &	   1.075    &    -3.892  \\
    3.3   &  2.688   &   -2.522 &	  0.9155    &    -4.384  \\
    3.5   &  2.629   &    -2.87 &	  0.7131    &    -4.916  \\
    3.7   &  2.567   &   -3.242 &	  0.4657    &    -5.485  \\
    3.9   &  2.491   &   -3.641 &	  0.1755    &    -6.118  \\
\hline                                  
\end{tabular}
\end{table}

\end{appendix}


\end{document}